\documentclass[12pt,aps,pre,preprint]{revtex4-1}   % style for Physical Review B and AJP are similar

\usepackage{amsmath}    % need for subequations
\usepackage{amsfonts}  % note how statements can be commented out
\usepackage{graphicx}   % for figures
\usepackage{bm}  %for bold in math mode
\usepackage{calligra}
\DeclareMathAlphabet{\mathcalligra}{T1}{calligra}{m}{n}
\DeclareFontShape{T1}{calligra}{m}{n}{<->s*[2.2]callig15}{}

\newcommand{\scripty}[1]{\ensuremath{\mathcalligra{#1}}}
\def\scrmag{\scripty{r}}

\def\scrhat{\hat{\scripty{r}}}
\def\scr2{\frac{\scrhat}{\scrmag^2}}

\begin{document}

%\title{Nonequilibrium Thermodynamics with Variable Kinetic Coefficients: a Case for Thermodynamic Induction} 
%\title{Thermodynamic Induction Resulting from Nonequilibrium Systems with Variable Kinetic Coefficients}
%\title{Nonequilibrium Systems with Variable Kinetic Coefficients and a Thermodynamic Induction Principle}
%\title{Thermodynamic Induction Exhibited in Nonequilibrium Systems with Variable Kinetic Coefficients}
\title{Thermodynamic Induction Effects Exhibited in Nonequilibrium Systems with Variable Kinetic Coefficients}

\author{S. N. Patitsas}
%\thanks{}
\affiliation{University of Lethbridge,\\4401 University Drive, Lethbridge AB, Canada, T1K3M4}


\date{\today}

\begin{abstract}

A nonequilibrium thermodynamic theory demonstrating an induction effect of a statistical nature is presented.  We have shown that this thermodynamic induction can arise in a class of systems that have variable kinetic coefficients (VKC).  In particular if a kinetic coefficient associated with a given thermodynamic variable depends on another thermodynamic variable then we have derived an expression that can predict the extent of the induction.  The amount of induction is shown to be proportional to the square of the driving force.  The nature of the intervariable coupling for the induction effect has similarities with the Onsager symmetry relations, though there is an important sign difference as well as the magnitudes not being equal.  Thermodynamic induction adds nonlinear terms that improve the stability of stationary states, at least within the VKC class of systems.  Induction also produces a term in the expression for the rate of entropy production that could be interpreted as self-organization.  Many of these results are also obtained using a variational approach, based on maximizing entropy production, in a certain sense.  Non-equilibrium quantities analogous to the free energies of equilibrium thermodynamics are introduced.  

\end{abstract}


\maketitle %\maketitle must follow title, authors, abstract and \pacs

\section{Introduction} \label{sec:intro}

In systems containing two or more irreversible transport mechanisms, reciprocal relations among transport coefficients have been observed for quite some time now.  Early examples include Kelvin's analysis of thermoelectric phenomena and Helmholtz's investigations into the conductivity of electrolytes.%\cite{Thomson1854,Helmholtz1876}

In 1931, Onsager published his seminal studies concerning the approach to thermodynamic equilibrium~\cite{Onsager1931,Onsager1931b}.  This is specifically a near-equilibrium, linear, theory where assumptions of local equilibrium apply.  Such a linearized approach naturally involves constant kinetic coefficients.  Onsager showed that there exists very general symmetries among such coefficients.  This work was further developed theoretically over the next couple of decades (see Refs.~\cite{Casimir1945,Callen1948,DeGrootMazur1954,DeGrootMazur1954b,vanKampen1954} for examples), while from the experimental side, many systems were studied in detail, including systems exhibiting particle diffusion, thermal conduction, electrical conduction, thermoelectricity, thermomagnetic, thermomechanical and galvanomagnetic effects, electrolytic transference, liquid helium fountain effects, and chemical reactions.  The experimental tests for the validity of the Onsager relations have been reviewed extensively by Miller~\cite{deGroot1947e,Miller1960}. In his review Miller points out that linearization is required in certain systems.  In particular, systems involving chemical reactions are often highly nonlinear and linear modeling is expected to be inaccurate.   The linear theory for approach to equilibrium was developed further when Prigogine studied stationary states and proved an important theorem on entropy production rates, i.e., the minimum entropy production principle~\cite{degroot, Jaynes1980}.

Examples of early attempts to deal with nonlinearities in nonequilibrium thermodynamics included nonlinear convection and accounting for relativistic effects~\cite{PrigogineGlansdorff1964,deGroot1968}.  Nonlinear chemical reaction thermodynamics was developed as well~\cite{Prigogine1965,Prigogine}.  This chemical work has continued on: for example with studies of cooperative electron transfer~\cite{Pohlmann1992,Tributsch2007}.  
Other examples of work related to Onsager relations and extension into nonlinear realms include a nonlinear analysis developed to describe nonelectrolyte transport in kidney tubules and a phenomenological approach to analyze nonlinear aspects of electrokinetic effects~\cite{Sauer1973,Gonzalez1994}.
%Gonzalez-Fernandez\cite{} used a nonlinear phenomenonological approach to analyze nonlinear aspects of electrokinetic effects.  
%He has 2 papers in j Noneq Therm in 83, 84.  check these. not yet.
%Chemical reactions often require nonlinear models for acccurate prediction.  Nonlinear nonequilibrium thermodynamics have been used to model cooperative electron transfer~\cite{Pohlmann1992,Tributsch2007}.
% did pick this up. Looks good.  -regarding  no experiments though.
%They followed this much later: \cite{Tributsch2007} with a real system. -interesting negentropy figure.

Further theoretical developments, partly aimed at extending the Onsager relations to the nonlinear domain, included a study of the time-reversal properties of Onsager's relations~\cite{Muschik1977}, as well as formulation of a generalized dissipation function~\cite{BatailleEdelinKestin1978}.  On the question of how to extend the Onsager relations, a clear answer has remained elusive.
%\cite{Muschik1977, BatailleEdelinKestin1978, BatailleKestin1979, Stockel1983}.  
%Several people worked in the nonlinear problem in the 1970s, including the Bataille, Edelin, Kestin group.  In they focus on chemical kinetics.  They review the nonlinear literature a little bit by dividing it into two proposals: Group 1 says symmetric matrix.  They refer to a number of publications but not listed!
%Kestin published a thermo book in 1968 and 1978.
%Proposal 2 is Ziegler.  -not checked yet (max entropy guy).
%In\cite{BatailleEdelinKestin1978} they claim to show that neither approach is valid in general.  Use this.  They go on to discuss a generalized dissipation potential.  
%In the 1980s, Stockel looked at generalizing Onsager's Relations to the nonlinear case, using a phenomenonological approach.  He likes focusing on time derivatives.
%We note that these approaches lack the simplicity of Prigogine's approach taken for linear dynamics; in particular his judicious choice to hold some variables constant.  Other variables are allowed to vary to find an extremum of some function, but not \textit{all} of them~\cite{Prigogine1965}.
%Drop any Discussion of equilibrium minimum or maximum entropy principles: \cite{Jaynes1978b} \cite{Jaynes1963} \cite{Jaynes1971} \cite{Jaynes1965} 
%max ent princ  \cite{Jaynes1980}\cite{Tykodi}  -not sure how to deal with these.  -but it's all linear
%Discussion of maximum entropy production principle: Zeigler   
In contrast to Prigogine's principle of minimum entropy production, Zeigler has proposed a principle of maximum entropy production.  This was based on studies of dissipation in thermomechanics and is a linear response theory~\cite{Zeigler1977}.  Since then, Zeigler's principle has stood in seemingly direct contradiction to Prigogine's principle, without clear resolution (see Ref.~\cite{Seleznev2006} for a recent review).  Bordel has discussed Zeigler's principle in the context of (linear) information theory~\cite{Bordel_2010}.
%Warren D. Smith has some objections 1999 to Onsager relations (linear) and mentions Pathria Stat mech 1972.  -seems like BS, no peer reviewed publication, just some crap on the internet
%Bordel is proponent of the max ent prod idea of Ziegler.\cite{Bordel_2010}  He has several examples: one is crystal growth.
%Kirkaldy: Rep Prog Phys 1992, can't find?  Lots of math here, but not convincing.
In this work we explore the possibility that this minimum vs. maximum controversy might be settled by looking at nonlinear systems approaching equilibrium, with the nonlinearity embedded into the kinetic coefficients. 

To help illustrate the question of what Onsager's reciprocal relations might look like in the nonlinear realm, we consider two thermodynamic variables, $a_1$ and $a_2$.  Then the customary equations for the linear nonequilibrium dynamics are~\cite{degroot, Reif, Pathria, LandauL}:
\begin{equation}
{\dot{a}}_1 = L_{11}X_1 +L_{12}X_2 ~, \label{a1L11L12}
\end{equation}
\begin{equation}
{\dot{a}}_2=  L_{21}X_1+ L_{22}X_2  ~. \label{a2L21L22}
\end{equation}
The famous result by Onsager stipulates that $L_{12}=L_{21}$.  Zero external magnetic fields are assumed throughout this work.  In order to emphasize our results reported here, we further suppose the off-diagonal Onsager coefficients are zero to linear order:
\begin{equation}
{\dot{a}}_1 = L_{11}X_1 ~,  \label{a1L11}
\end{equation}
\begin{equation}
{\dot{a}}_2=   L_{22}X_2 ~.  \label{a2L22}
\end{equation}
Up to linear order these two variables are now uncoupled.  If we next allow for the $L_{11}$ kinetic coefficient to have a functional dependence on variable $a_2$ then extra nonlinear terms will be generated.  If $a_2$ happens to be the same as $X_2$ within a constant, then by keeping only the next leading order term, and making the direct substitution: $L_{11}\rightarrow L_{11}+cX_2$, we may modify Eq.~(\ref{a1L11}) to look like:
\begin{equation}
{\dot{a}}_1 = L_{11}X_1 + (c X_1) X_2  ~,\label{claim1}
\end{equation}
where $L_{11}$ is now strictly evaluated at $X_2=0$.  The $c X_1 X_2$ term is what we refer to as the \textit{direct} contribution to nonlinear dynamics.  One may then ask whether Eq.~(\ref{a2L22}) also gets modified.  Put another way, can the dependence of $L_{11}$ on $a_2$ indirectly induce an extra term into Eq.~(\ref{a2L22}) which would change the way $a_2$ changes with time?  Naive application of the Onsager symmetry would suggest $L_{12}=cX_1$, adding the term $(c X_1) X_1$ to Eq.~(\ref{a2L22}).  However there is no known theoretical justification for this, since this theory is inherently nonlinear and the proofs justifying the Onsager symmetry are not valid.  We will show below that adding the term $c X_1^2$ is on the right track, but the magnitude will be smaller and more importantly there is an added minus sign.  The actual equation for ${\dot{a}}_2$ becomes:
\begin{equation}
{\dot{a}}_2=  - (r c X_1) X_1 + L_{22}X_2  ~, \label{claim2}
\end{equation}
where $0<r<1$.  We show in this work that this result is valid in the case where the characteristic timescale for $a_2$ is fast compared to that of $a_1$.  We refer to the new term in these equations as the \textit{thermodynamic induction} term.  This important result will be presented as the first of three theorems in total.

After introducing the thermodynamic induction concept in Sec.~\ref{sec:Langevin} with a specific example, we develop, in Sec.~\ref{sec:gentheory}, a more general theory for thermodynamic induction involving $n$ variables, and using a classical statistical mechanical approach following that of Onsager.  We present results for a class of nonlinear systems we refer to as the variable kinetic coefficients (VKC) class.  The nonlinearity of these systems arises naturally: systems exhibiting behavior with temperature dependent thermal conductivity, electrical conductivities of solutions that depend on solute concentration, diffusion coefficients and chemical reaction rate coefficients that depend on temperature, are but a few examples that can be treated with this theory.  
%The key feature is prominent use of the factor $\exp(\Delta S/k_B)$.  
%This factor is used extensively when describing approaches to equilibrium.  Recently it has been used in formulating the Searles-Evans Fluctuation Theorem. ~ 
In Sec.~\ref{statstates} we analyze the ramifications of thermodynamic induction by studying quasistationary states.  These states prove to be very valuable and play a key role in our second theorem: a dynamical nonequilibrium result that resembles Le Chatelier's principle of equilibrium thermodynamics.  This leads to the final theorem presented in Sec.~\ref{sec:maxent} as a principle of maximum free entropy production.
% new and interesting principle brought about by an all important minus sign: 
Before concluding, we present some example calculations in Secs.~\ref{sec:2var}, \ref{sec:numer}, and~\ref{sec:thermal}.

\section{Langevin particle and the dynamical reservoir} \label{sec:Langevin}

%Here we follow Reif\cite{Reif15} very closely, pgs 568-569.  
We consider and, at first, briefly review the classical diffusing particle with dynamics governed by the Langevin equation in one dimension~\cite{Reif15}.  Assuming no external forces aside from the quickly varying random force $F(t)$ the Langevin equation is
\begin{equation}
m\frac{dv}{dt}= +F(t)~.
\end{equation}
The force $F(t)$ varies on a timescale $\tau^*$.  To integrate this equation we must take a timestep much larger than $\tau^*$.  The average value of $v$ will change much more slowly than
$F(t)$, so we take care to choose a timestep $\Delta t$ which is large enough so that $\Delta t\gg\tau^*$, but small enough that the change in $\bar{v}$ is small.  The result is
\begin{equation}
m\langle[v(t+\Delta t)-v(t)]\rangle= \int_t^{t+\Delta t}\langle F(t')\rangle dt' \label{DeltavEqn}
\end{equation}
where the ensemble average has been taken on both sides.  These ensemble averages are not over equilibrium states since $\langle F\rangle_0 =0$.  The states of the reservoir environment (heat bath) respond quickly to the particle so when the mean velocity $\bar{v}$ is nonzero, the environment responds and thus changes $\langle F\rangle$.  Since the response of the environment is fast, local equilibrium conditions apply.  We invoke the factor $\exp (\Delta S/k_B)$ to describe deviations from the equilibrium and convert the averaging procedure to evaluating equilibrium averages i.e.
\begin{equation}
\langle F(t')\rangle = \langle e^{\frac{\Delta S}{k_B}} F(t')\rangle_0 \approx \langle  (1+\Delta S/k_B) F(t')\rangle_0 =\frac{1}{k_B} \langle (\Delta S) F(t')\rangle_0 \label{Ftprime}
\end{equation}
where $\Delta S$ represents the change in the reservoir entropy due to its reaction to the change in average velocity.   If the reservoir temperature is $T$ then for the standard (linear) component to $\Delta S$ one obtains $T \Delta S_{lin}=\Delta E=-\int_t^{t'}v(t'')F(t'')dt''$.  This leads to the result $\langle F_{lin}(t')\rangle|_0 =-\frac{1}{k_B T}\bar{v}(t)\int_t^{t'}dt''K(s)$ where  $K(s)=\langle F(t')F(t'+s)\rangle|_0$ and $s\equiv t''-t'$.  Thus to linear order:
\begin{equation}
m\langle[v(t+\Delta t)-v(t)]\rangle= -m\beta(\Delta t) \bar{v}(t)
\end{equation}
where $\beta=\frac{1}{2m k_B T}\int_{-\infty}^{\infty}K(s)ds$.  Using the form $K(s)=K(0)\exp(-|s|/\tau^*)$ we obtain the useful relation: 
%$\beta=\frac{K(0)\tau^*}{m k_B T}$
\begin{equation}
m k_B T\beta= \tau^* K(0).  \label{K0tau*}
\end{equation}

\subsection{Dynamical reservoir}

So far we have considered our particle interacting with a standard heat reservoir in equilibrium as a whole.  We introduce the \textit{dynamical reservoir} as a reservoir that is out of equilibrium and in the process of slowly returning to equilibrium.  This approach is governed by a set of kinetic coefficients.  In general these coefficients are constant to linear order but in real systems may depend somewhat on other thermodynamic variables.  

Here we consider what may be the somewhat artificial, yet instructive, case where one dynamical reservoir kinetic coefficient depends on the velocity of the particle considered here.  The reservoir dynamics is then governed by the equation, applicable when the system is not too far from equilibrium:
\begin{equation}
\dot{a}=M X
\end{equation}	
where $a$ is the small deviation from equilibrium of the reservoir variable, $X$ is the conjugate force, and $M=L+\gamma v$ is the kinetic coefficient.  The coefficient is assumed to depend on $v$ in a linear fashion, where $L$ is constant and $\gamma$ also a (small) constant.  This dependance creates nonlinear effects and is what couples our particle to the dynamical reservoir.  The approach to equilibrium creates entropy at a rate $\sigma=M X^2$ so that in a time interval, $dt$, an amount of entropy $\sigma dt$ is created.  The linear component of this entropy change, $LX^2dt$, will occur regardless of the state of the particle and so will have no effect on the particle dynamics.  As we will see, the nonlinear component will have a significant effect.  We consider the nonlinear part of the entropy change 
\begin{equation}
\Delta S_{nonlin}(t')=\int_t^{t'}dt'' \gamma X(t'')^2 v(t'') \approx \gamma v X^2\int_t^{t'}dt''v(t'')~,
\end{equation}
where we have pulled the slowly varying function $X(t)$ out of the integral, and inserted into Eq.~(\ref{Ftprime}) to obtain:
\begin{equation}
\langle F(t')\rangle|_{nonlin} =\frac{1}{k_B} \langle [\Delta S_{nonlin}(t')] F(t')\rangle_0 =\frac{\gamma X^2}{k_B} \int_t^{t'}dt'' \langle (v(t'') F(t')\rangle_0  ~.
\end{equation}
We convert to the force-force correlation function as:
\begin{equation}
\langle v(t'') F(t')\rangle_0 = \int_{-\infty}^{t''}dt'''\langle \dot{v}(t''') F(t')\rangle_0  =\frac{1}{m} \int_{-\infty}^{t''}dt''' K(t'-t''') ~.
\end{equation}
Using Eq.~(\ref{K0tau*}) we find
\begin{equation}
\langle F(t')\rangle|_{nonlin} = \gamma X^2 \beta T \tau^* [ 1 - e^{-(t'-t)/\tau^*} ] ~,\label{Ftanswer}
\end{equation}
and assuming $\Delta t\gg\tau^*$ we find:
\begin{equation}
\int_t^{t+\Delta t}dt'\langle F(t')\rangle|_{nonlin} =\gamma X^2 \tau^* T\beta (\Delta t)~.
\end{equation}

Putting together the linear and nonlinear terms:
\begin{equation}
m\langle[v(t+\Delta t)-v(t)]\rangle= -m\beta \bar{v}(t) (\Delta t) + \gamma X^2 T \beta\tau^* (\Delta t) ~,
\end{equation}
%\begin{equation}
%m\langle[v(t+\tau)-v(t)]\rangle= -m\beta\tau\bar{v}(t) + \gamma X^2\tau T \beta\tau
%\end{equation}
or
\begin{equation}
m\langle\dot{v}(t)\rangle= -\beta m\bar{v}(t) + \gamma X^2 T\beta\tau^*~.
\end{equation}
We see as the result an extra force term, constantly pushing the particle toward greater (lesser) velocity if $\gamma$ is positive (negative).  This is an example of what we in general call \textit{thermodynamic induction}.  Fluctuations play an important role, as there is no direct external force on the particle.  Instead, by having the particle randomly accessing states, there is a statistical bias towards those particle states that allow the dynamical reservoir to create entropy at a greater rate.  The induced force is small, containing factors $\gamma$, $X$, and $\tau^*$, all of which are assumed to be small in some sense. Though this example is somewhat contrived, it does make the point of making a case for the induction effect for a very well known system.  In order to discover more plausible physical examples demonstrating this interesting induction effect we proceed to a more general thermodynamic analysis. 

\section{General Theory for Thermodynamic Induction Effect}  \label{sec:gentheory}

Consider an isolated system described by $n$ thermodynamic variables $x_i$ with equilibrium values $x_{i_0}$.  In this work we follow closely the approach presented in Ref.~\cite{degroot}, in particular by considering only discrete variables, and leaving the treatment of continuous variables for future work.  The discrete approach taken here suffices to clearly demonstrate the thermodynamic induction effect.  In fact even two variables will suffice, as we will show below.  

The total entropy $S_T$ may be written as a function of the $n$ variables $x_i$.  The change $\Delta S_T$, from equilibrium, of the total entropy, may be written as a function of the $n$ state variables $a_i=x_i-x_{i_0}$.  Generalized forces (affinities) are defined by $X_i=\frac{\partial \Delta S_T}{\partial a_i}$.  These conjugate parameters are interrelated as:    
\begin{equation}
X_i=-\sum_{j=1}^n g_{ij}a_j         ~,    \label{Xga}
\end{equation} 
and
\begin{equation}
a_i=-\sum_{j=1}^n g_{ij}^{-1}X_j       ~  ,    \label{agX}
\end{equation} 
where $g_{ij}=g_{ji}$, i.e. the $g$ matrix is symmetric, and with the same symmetry holding for the inverse matrix.  In order to ensure thermodynamic stability, both the $g$ matrix and its inverse must be positive definite.  These relations place the $X_i$ variables on equal footing with the $a_i$ variables.  
%Specifying all the $X_i$ suffices to describe the thermodynamic state.  
We note that these forces are defined as differential quantities, and that they are not the same as $\frac{\partial S_T}{\partial a_i}$.  See Ref.~[\onlinecite{Patitsas2013}] for an example worked out with two variables, i.e., $n=2$.  For each variable $a_k$ we are to think of a transfer of some quantity from one region to another.  The entropy of one region decreases while the entropy of the other increases by a yet slightly larger amount, with net increase $\Delta S_{k}$.  Though many of these transfer processes may share the same regions of space, there may in all be as many as $2n$ distinct regions.
%In equilibrium certain mean values of variable combinations are readily determined~\cite{degroot}.  One that we will have a particular use for is:
%\begin{equation}
%\left\langle a_i(t) a_j(t)\right\rangle_0=k_B g_{ij}^{-1}~.   \label{aijavg}
%\end{equation}

When considering relaxation towards equilibrium, the customary approach is to write dynamical equations that relate the time derivatives $\dot{a}_i$ to linear combinations of the forces $X_j$.  The kinetic coefficients are closely related to transport coefficients such as thermal conductivity, electrical conductivity, etc.  These transport processes are irreversible:  the internal entropy increases with time as the system approaches equilibrium.

Before analyzing the induction effect, which is inherently nonlinear, we briefly review the general approach.  Since the variables treated here are statistical, a dynamical approach should use coarse-grained averages of the time derivatives.  For variable $x_i$ the timestep $\Delta t_i$ chosen for the dynamics must be much larger than the relevant force (fluctuation) correlation time $\tau_i^*$, which is typically very short.  %This timescale $\tau_i$ will be discussed in more detail below. 
We closely follow the notation in Ref.~\onlinecite{degroot}, and use a bar to denote the coarse-grained time derivative as $\bar{\dot{a}}_i$.  The definition is
\begin{equation}  
\bar{\dot{a}}_i\equiv\frac{1}{\Delta t_i}\int_t^{t+\Delta t_i}{\langle\dot{a}_i\rangle dt'}=\frac{1}{\Delta t_i}\langle a_i(t+\Delta t_i)-a_i(t)\rangle  ~  .
\end{equation}
We note that the symbols $\langle~\rangle$ denote ensemble averages over accessible microstates of the system.  Addition of the subscript, 0, will refer to equilibrium states in particular.

In order to evaluate this coarse-grained derivative we closely follow the approach in Ref.~\cite{Reif15}.  In this approach, a nonequilibrium average value is determined by evaluating an equilibrium average accompanied by the insertion of the ubiquitous $\exp(\Delta S_T/k_B)$ probability weighting factor~\cite{LandauL,degroot}.  We emphasize that this must be the total system entropy change.  This weighting factor has played an important role in thermodynamics for a long time and continues to receive attention.  For example, the factor has been firmly established recently as a consequence of the fluctuation theorem of Evans and Searles~\cite{Searles2002}.    The theorem is derived from exact, detailed microscopic nonequilibrium dynamics, and the $\exp(\Delta S_T/k_B)$ factor is recovered in the short-time limit of the theorem.
Explicitly then:  
\begin{equation}
\langle\dot{a}_i(t')\rangle=\left\langle\dot{a}_i(t')e^{\frac{\Delta S_T(t'-t)}{k_B}}\right\rangle_0~,   \label{expS}          
\end{equation}
%where we consider a timestep $\tau'<\tau$, $\tau'>>\tau^*$, such that $t'=t+\tau'$, and 
where $\Delta S_T(t'-t)\equiv S_T(t')-S_T(t)$.  Since the change in entropy is small, a linear expansion is warranted, leaving:
\begin{equation}
\langle\dot{a}_i(t')\rangle=\frac{1}{k_B}{\langle\dot{a}_i(t')\Delta S_T(t'-t)}\rangle_0~,   \label{adot2b}
\end{equation}
since $\langle \dot{a}_i \rangle_0=0$.  Integrating both sides of Eq.~(\ref{adot2b}) over the time interval $\Delta t_i$ gives the coarse-grained time derivative:
\begin{equation}
\bar{\dot{a}}_{i}=\frac{1}{\Delta t_i}\int_t^{t+\Delta t_i}\langle\dot{a}_i(t')\rangle dt'=\frac{1}{k_B\Delta t_i}\int_{t}^{t+\Delta t_i}{dt'\langle\dot{a}_i(t')\Delta S_T(t'-t)\rangle_0}~. \label{adot3}
\end{equation} 
Making use of the generalized forces, the entropy change is
\begin{equation}
\Delta S_T=\sum_{j=1}^n X_j \Delta a_j    ~. \label{XjDeltaAj}
\end{equation}
In terms of the rate of entropy production, $\sigma_T=\frac{dS_T}{dt}$, we note that
\begin{equation}
\sigma_T=\sum_{j=1}^n X_j \dot{a}_j    ~,
\end{equation}
so we may express $\Delta S_T$ as
\begin{equation}
\Delta S_T=\int_{t}^{t+\Delta t_i}{\sigma dt''}=\sum_{j=1}^n \int_{t}^{t+\Delta t_i}{ X_j \dot{a}_j dt''}    ~. \label{DeltaSint}
\end{equation}

\subsection{Review of linear case}

Substituting Eq.~(\ref{DeltaSint}) into Eq.~(\ref{adot3}) leaves
\begin{equation}
\bar{\dot{a}}_i=\frac{1}{k_B\Delta t_i}\int_{t}^{t+\Delta t_i}{dt' \int_{t}^{t+\Delta t_i}dt''\langle\dot{a}_i(t')\sum_{j=1}^n X_j \dot{a}_j(t'')\rangle_0}~. 
\end{equation}
Closely following Ref.~\cite{Reif15} by taking the slowly varying force functions $X_j$ out of the integrals, and with a few more elementary steps one obtains
\begin{equation}
\bar{\dot{a}}_i=\sum_{j=1}^n L_{ij}X_j     \label{Lij} ~,
\end{equation}
where 
\begin{equation}
L_{ij}=\frac{1}{k_B}\int_{-\infty}^0 ds K_{ij}(s)~,  \label{LijKij}
\end{equation}
and $K_{ij}$ are the cross-correlation functions defined by $K_{ij}\equiv\langle \dot{a}_i(t)\dot{a}_j(t+s)\rangle_0$.  The Onsager reciprocal symmetry $L_{ij}=L_{ji}$ is obtained by using time-reversal symmetry and assumption of zero external magnetic field~\cite{Reif15}.  In the linear regime, the kinetic coefficients $L_{ij}$ are constants.\footnote{See Ref.~[\onlinecite{Patitsas2013}] for an example with this symmetry explicitly verified inside of a specific model, i.e., classical ideal gas effusion.}

One actually solves for the system dynamics by combining Eqs.~(\ref{agX}),~(\ref{Lij}) to obtain
\begin{equation}
{\dot{X}}_i=-\sum_{j=1}^n A_{ij}X_j     \label{Xlindyneq} ~,
\end{equation}
where $A_{ij}=\sum_{k=1}^n g_{ik} L_{kj}$.  The eigenvalues of the $A$ matrix have units of $s^{-1}$ and these define the $n$ (relaxation) timescales $\tau_i$ for the system.  In the special case where the matrices $g_{kl}$ and $A_{kl}$ are diagonal, then we have the following relation:
\begin{equation}
g_{kk}L_{kk}\tau_k=1     \label{gLtauOne} ~.
\end{equation}
The rate of total entropy production is 
\begin{equation}
\sigma_T =\sum_{i=1}^n \sum_{j=1}^n L_{ij}X_j X_i   ~.               \label{sigma}
\end{equation}

\subsection{Nonlinear case}

Next we consider the case where the kinetic coefficients are not constants but instead depend on the variables $\{x_i\}$.  We assume these dependencies to be weak so that the nonlinear terms generated are small.  This restricts our analysis to a subset of all possible systems which we term VKC.  We understand that there may be systems that are nonlinear and yet do not fall into this VKC class.

Our aim then is to adapt the basic approach used in the linear analysis by adding in these small nonlinear terms.  The new (variable) coefficients will be labeled $M_{ij}(\{a_l\})$.  In equilibrium, where $a_l=0$, they take the (constant) values $L_{ij}$ i.e.
\begin{equation}
M_{ij}(\{a_l=0\})=L_{ij}~.               
\end{equation}
The Onsager symmetry 
\begin{equation}
M_{ij}=M_{ji} \label{MijMji}
\end{equation}
 still holds.  One can follow the proof given in the previous section for $L_{ij}=L_{ji}$ still holds even when these coefficients depend on another thermodynamic variable that is not $a_i$ or $a_j$.  Use of the coefficients $M_{kl}$ will make the dynamical equations nonlinear.  Our task then is to determine how Eq.~(\ref{Lij}) is to be modified. 

\subsubsection{Key assumptions}

We make four assumptions before proving Theorem 1, our main result.

\textbf{Assumption 1}

We make linear expansions in the variables $\{a_l\}$, and ignore higher order terms.  This defines a new set of constants $\gamma_{ij,l}$ such that
\begin{equation}
M_{ij}=L_{ij}+\sum_{l}\gamma_{ij,l} a_l~,       \label{gamma}               
\end{equation}
and
\begin{equation}
\gamma_{ij,l}=\left(\frac{\partial M_{ij}}{\partial a_l}\right)_{a_l=0}   ~.\label{gammaijk}
\end{equation}
The Onsager symmetry condition Eq.~(\ref{MijMji}) implies the following similar symmetry:
\begin{equation}
\gamma_{ij,l}=\gamma_{ji,l}   ~.
\end{equation}
The $\gamma_{ij,l}$ coefficients describe the variability of the kinetic coefficients, to leading order.  This is the most reasonable approach to account for the variability of kinetic coefficients.  We assume that the coefficient $\gamma_{ij,k}$ for the linear order corrections are not especially small somehow.  In such a case we may have to treat correction terms up to quadratic order in $a_k$.  

\textbf{Assumption 2}

Our second assumption is that of the $n$ variables in question, $m$ are of the slow variety while $n-m$ are quickly varying in time.  The slow variables are labeled with indices $i,j$ with index values ranging from 1 to $m$, and the fast variables are labeled with indices $k,l$ with these index values ranging from $m+1$ to $n$.   More exactly the assumption is $\tau_k\ll\tau_i$ for all $k>m$ and $i\leq m$.  This is quite a restrictive condition.  Fast and slow variables cannot be coupled to linear order.  This means that both the $g_{pq}$ and $L_{pq}$ matrices must both have block-diagonal forms with two main diagonal blocks, slow and fast, i.e., $g_{ik}=0$, $L_{ik}=0$ when $i>m$ and $k\leq m$.   The slow variables could represent the dynamical reservoir, though we point out that the slow systems need not be large in any sense.  It is only their large characteristic timescales that matter here.

\textbf{Assumption 3}

Our third assumption is that of no variability in the kinetic coefficients $M_{kl}$ associated with fast variables, i.e., $\gamma_{kl,q}=0$ for all $k,l>m$.  We note that this assumption means the fast diagonal block part of the $L$ matrix is unchanged.  Since $g_{kl}$ and $L_{kl}$ are both symmetric and positive definite, we may find a transformation to new variables for which both  $g_{kl}$ and $L_{kl}$ are diagonal.  This may involve both rotation and stretching transformations, similar to the procedure used to solve for normal modes in a system of many masses and springs~\cite{Symon}.  Here, we will assume that these transformations have already been accomplished for the fast variables.  Though the physical interpretation of these transformed fast variables may be difficult, the transformation helps in the following analysis, as these variables are decoupled from each other.  Thus, with no loss of generality we take the fast block of both matrices $g_{kl}$ and $L_{kl}$ as diagonal.  

\textbf{Assumption 4}

Our fourth assumption is that the $M_{ij}$ coefficients (for the slow variables) depend only on fast variables, i.e., $\gamma_{ij,q}=0$ for $q\leq m$.   We note that this assumption means the slow diagonal block part of the $L$ matrix is unchanged.  This last assumption is mostly for tidying up the algebra.  In the case of only one slow variable it makes no difference.  In this case the one slow variable $a_1$ would solely play the role of dynamical reservoir.

\subsubsection{Direct contributions}

In order to determine how Eqs.~(\ref{Lij}) are to be modified, we treat the slow and fast variables separately.  In so doing, we make a clear division between what we term direct and induced contributions.  Both types of contributions are essential and also suffice to describe the system to leading order of nonlinearity. 

The direct nonlinear contribution arises from simply substituting in $M_{ij}$ for $L_{ij}$ in Eq.~(\ref{Lij}), while focusing on the dynamics for the slow variables ($i\leq m$): 
% From here on we will drop the overbar denoting coarse-grained time derivatives. (FIX THIS)  Explicitly:     
\begin{equation}
\left(\dot{a}_i\right)_{dir}=\sum_{j=1}^m  X_j \sum_{k=m+1}^{n}\gamma_{ij,k}a_k       ~.       
\end{equation}
%where it is understood that in this equation, $i\leq m$ i.e. applicable to slow variables. %k is fast.  What about j?  It sure helps if j is slow only.  
Making use of Eq.~(\ref{agX}), 
%\begin{equation}
%\left(\dot{a}_i\right)_{dir}=-\sum_{j=1}^m  X_j \sum_{l=m+1}^{n}\gamma_{ij,l}\sum_{k=m+1}^n g_{lk}^{-1}X_k      ~,       
%\end{equation}
\begin{equation}
\left(\dot{a}_i\right)_{dir}=-\sum_{j=1}^m  X_j \sum_{k=m+1}^{n}\gamma_{ij,k} g_{kk}^{-1}X_k      ~,       
\end{equation}
or,
\begin{equation}
\left(\dot{a}_i\right)_{dir}=\sum_{k=m+1}^n N_{ik} X_k      ~, \label{adirectX}      
\end{equation}
where
%\begin{equation}
%N_{ik}=-\sum_{j=1}^m \sum_{l=m+1}^{n}\gamma_{ij,l} g_{lk}^{-1}X_j~,~~~~i\leq m,~k>m      ~.    \label{Ndir}  
%\end{equation}
\begin{equation}
N_{ik}=-g_{kk}^{-1}\sum_{j=1}^m \gamma_{ij,l} X_j =-L_{kk} \tau_{k}\sum_{j=1}^m \gamma_{ij,l} X_j~,~~~~i\leq m,~k>m      ~,    \label{Ndir}  
\end{equation}
having made use of Eq.~(\ref{gLtauOne}).  The coefficients $N_{ik}$ are not constant because of the nonlinearity.  We note that these coefficients depend only on slowly varying forces $X_j$.
We also note that the direct contributions do not change the dynamics for the fast variables.

\subsubsection{Induced contributions}

We show that the fast variables are also affected, indirectly.  Modification of the dynamical equations for fast variables gives rise to effects we designate as thermodynamic induction.   The induced contributions will play a key role in the fast variable dynamics.  We formulate the following theorem for this induction effect.

\textbf{Theorem 1 (principle of thermodynamic induction)}

\begin{equation}
\left(\bar{\dot{a}}_k\right)_{ind}=\sum_{i=1}^{m}N_{ki} X_i, \label{aindX}
\end{equation}
where
%\begin{equation}
%N_{ki}=\sum_{j=1}^{m}\sum_{l=m+1}^{n}\gamma_{ij,l} g_{kl}^{-1} X_j ~,~~~~k>m ,~i\leq m~.  \label{Nind}
%\end{equation}
\begin{equation}
N_{ki}= L_{kk} \tau^*_{k} \sum_{j=1}^{m}\gamma_{ij,k}   X_j,~~~~k>m ,~i\leq m~.  \label{Nind}
\end{equation}
%This result constitutes the principle of thermodynamic induction.

The key step in our proof is to revisit Eqs.~(\ref{adot3}) and make a distinction between linear and nonlinear contributions to $\Delta S_T$:
\begin{equation}
\Delta S_T=\Delta S_{lin}+\Delta S_{nonlin}   ~ .
\end{equation}
The treatment for $\Delta S_{lin}$ is just as described above and leads to Eq.~(\ref{Lij}).  For the nonlinear treatment we again start from the expression for the rate of entropy production:
\begin{equation}
\Delta S_{nonlin}(t')=\int_t^{t'}\sigma_{nonlin} dt''    .  \label{DeltaS}
\end{equation}
Into this equation we insert Eq.~(\ref{sigma}) with the substitution $L_{ij}\rightarrow M_{ij}$.  Using Eq.~(\ref{gamma}) and extracting the nonlinear terms gives
\begin{equation}
\sigma_{nonlin}=\sum_{i=1}^{m}\sum_{j=1}^{m}\sum_{l=m+1}^{n}\gamma_{ij,l} X_i X_j a_l ~,    \label{signonlin}
\end{equation}
where we have made use of Assumptions 3 and 4.  Substituting Eq.~(\ref{signonlin}) into Eqs.~(\ref{adot3}) and~(\ref{DeltaS}) gives:
\begin{equation}
\left(\bar{\dot{a}}_k\right)_{ind}=\frac{1}{k_B\Delta t_k}\int_{t}^{t+\Delta t_k}dt'\int_t^{t'}dt'' \langle\dot{a}_k(t')\sum_{i=1}^{m}\sum_{j=1}^{m}\sum_{l=m+1}^{n}\gamma_{ij,l} a_l(t'') X_i X_j\rangle_0~,
\end{equation}
where the coarse-graining timescale $\Delta t_k\gg\tau_k^*$. 
%Our next step is to derive analytic expressions that describe the emount of entropic induction.
%Save for conclusions:
%We also note here that the two assumptions just made are not fundamentally necessary; The induction effect should exist in some systems not satisfying these assumptions.  We simply make these assumptions to simplify the analysis in order to arrive at analytic expressions that describe to what degree a given system might be affected by the induction effect.
%Proceeding with the analysis, we first consider the temporal variation of $a_k$ for only the fast variables i.e. $k>m$.  
Pulling out the slowly varying $X_i$, $X_j$ functions %~\footnote{Here indices $i,j$ will represent slow variables and $k,l$ fast variables.},
\footnote{This procedure easily allows for the possibility of the parameters $\gamma_{kl,q}$ depending on the slow parameters $\{a_i\}$, though here we will assume these are strictly constant.}:
\begin{equation}
\left(\bar{\dot{a}}_k\right)_{ind}=\frac{1}{k_B\Delta t_k}\sum_{i=1}^{m}X_i\sum_{j=1}^{m}X_j\sum_{l=m+1}^{n}\gamma_{ji,l}\int_{t}^{t+\Delta t_k}dt'\int_t^{t'}dt''\langle\dot{a}_k(t') a_l(t'') \rangle_0~. 
\end{equation}
Proceeding similarly as in Sec.~\ref{sec:Langevin}:
\begin{equation}
\left(\bar{\dot{a}}_k\right)_{ind}=\frac{1}{k_B\Delta t_k}\sum_{i=1}^{m}X_i\sum_{j=1}^{m}X_j\sum_{l=m+1}^{n}\gamma_{ji,l}\int_{t}^{t+\Delta t_k}dt'\int_t^{t'}dt'' \int_{-\infty}^{t''}dt'''\langle\dot{a}_k(t') \dot{a}_l(t''') \rangle_0~. 
\end{equation}
Noting that $\langle\dot{a}_k(t') \dot{a}_l(t''')\rangle_0= K_{kl}(t'-t''')$ is the correlation function and making use of Eq.~(\ref{LijKij}), we see that with the fast block part of the $M$ matrix being diagonal, that
%\begin{equation}
%\left(\bar{\dot{a}}_k\right)_{ind}=\frac{1}{k_B\Delta t_k}\sum_{i=1}^{m}X_i\sum_{j=1}^{m}X_j\sum_{l=m+1}^{n}\gamma_{ji,l}\int_{t}^{t+\Delta t_k}dt'\int_t^{t'}dt'' \int_{-\infty}^{t''}dt''' K_{kl}(t'-t''')  
%\end{equation}
$K_{kl}=0$ when $k\neq l$, so
\begin{equation}
\left(\bar{\dot{a}}_k\right)_{ind}=\frac{1}{k_B\Delta t_k}\sum_{i=1}^{m}X_i\sum_{j=1}^{m}X_j\gamma_{ji,k}\int_{t}^{t+\Delta t_k}dt'\int_t^{t'}dt'' \int_{-\infty}^{t''}dt''' K_{kk}(t'-t''')  ~.
\end{equation}
%\begin{equation}
%\left(\bar{\dot{a}}_k\right)_{ind}=\frac{1}{k_B}\sum_{i=1}^{m}X_i\sum_{j=1}^{m}X_j\sum_{l=m+1}^{n}\gamma_{ji,l} k_B L_{kl} \tau^*_{kl}  
%\end{equation}
With the assumptions $\Delta t_k\gg\tau_k^*$ we find
\begin{equation}
\left(\bar{\dot{a}}_k\right)_{ind}=\sum_{i=1}^{m}X_i\sum_{j=1}^{m}X_j\gamma_{ji,k} L_{kk} \tau^*_{k} ~, \label{ainfexplicit}
\end{equation}
and we have our result i.e. Eqs. (\ref{aindX}) and (\ref{Nind}).
%After some elementary steps (See Ref.~\onlinecite{Reif15}) we obtain
%\begin{equation}
%\left(\bar{\dot{a}}_k\right)_{ind}=\frac{1}{k_B}\sum_{i=1}^{m}X_i\sum_{j=1}^{m}X_j\sum_{l=m+1}^{n}\gamma_{ji,l}\langle a_k(t) a_l(t) \rangle_0~. 
%\end{equation}
%Use of Eq.~(\ref{gammaijk}) gives the desired form for the induction terms:
%\begin{equation}
%\left(\bar{\dot{a}}_k\right)_{ind}=\sum_{i=1}^{m}\sum_{j=1}^{m}\sum_{l=m+1}^{n}\gamma_{ij,l} g_{kl}^{-1} X_i X_j ~,  \label{ainfexplicit}
%\end{equation}
%\begin{equation}
%\left(\bar{\dot{a}}_k\right)_{ind}=\sum_{i=1}^{m}N_{ki} X_i ~, \label{aindX}
%\end{equation}
%where
%\begin{equation}
%N_{ki}=\sum_{j=1}^{m}\sum_{l=m+1}^{n}\gamma_{ij,l} g_{kl}^{-1} X_j ~,~~~~k>m ,~i\leq m~.  \label{Nind}
%\end{equation}
%\begin{equation}
%N_{ki}= L_{kk} \tau^*_{k} \sum_{j=1}^{m}\gamma_{ij,k}   X_j ~,~~~~k>m ,~i\leq m~.  \label{Nind}
%\end{equation}
Defining the dimensionless ratios $r_k$ as 
\begin{equation}
r_k\equiv \frac{\tau_k^*}{\tau_k} ~,
\end{equation}
allows us to express the thermodynamic induction theorem in the following form:
\begin{equation}
N_{ki}=-r_k N_{ik} ~. \label{relateIndDir}
\end{equation}
The induced terms are of opposite sign than the corresponding direct terms.  They are also smaller in magnitude since we have assumed $r_k\ll 1$. 

\subsubsection{Dynamical equations}

Equations~(\ref{Lij}),~(\ref{adirectX}),~(\ref{aindX}) combine to give the following dynamical equations:
\begin{equation}
\bar{\dot{a}}_p=\sum_{q=1}^n (L_{pq}+N_{pq})X_q  ~,   \label{Neqs} 
\end{equation}
with the nonlinear kinetic coefficients $N_{pq}$ specified by Eqs.~(\ref{Ndir}),~(\ref{Nind}), directly in terms of forces $X_i$.  We note that all of the nonlinear terms reside in the off-diagonal blocks of the $N_{pq}$ matrix, direct terms in the upper-right block, induced terms in the lower-left block.  Insertion of Eq.~(\ref{agX}) gives $n$ coupled nonlinear first order differential equations as:
\begin{equation}
-\sum_{q=1}^n g_{pq}^{-1} X_q=\sum_{q=1}^n (L_{pq}+N_{pq})X_q  ~.   \label{NeqsJustX} 
\end{equation}
These equations could be used to solve for the transient response after the system suffers a large fluctuation.  The procedure would involve solving these equations subject to a set of initial conditions $X_i(0)$.  Thinking of $L_{pq}+N_{pq}$ as a matrix we see that the diagonal blocks do not change at this order of analysis.  The off-diagonal blocks contain nonconstant elements which create nonlinear dynamics.  We do note that the variations in time for these kinetic coefficients will be slow.  These coefficients will appear to be almost constant as far as the fast variables are concerned.  
% Finally, the matrix is antisymmetric which stands in contrast to the symmetric matrix for linear dynamics, $L$.

\subsubsection{Entropy production}

For each variable we may define
\begin{equation}
\sigma_{p}\equiv \dot{a}_p X_p ~.
\end{equation}
Adding up these terms for slow variables only gives
\begin{equation}
\sigma_{slow}\equiv\sum_{i=1}^m \dot{a}_i X_i ~.\label{sigslow}
\end{equation}
For fast variables:
\begin{equation}
\sigma_{fast}\equiv\sum_{k=m+1}^n \dot{a}_k X_k \label{sigfast}
\end{equation}
so that $\sigma_T=\sigma_{slow}+\sigma_{fast}$. Using Eqs.~(\ref{sigma}),~(\ref{adirectX}),~(\ref{aindX}),
\begin{equation}
\sigma_{slow}= \sum_{i=1}^m \sum_{j=1}^m L_{ij}X_i X_j + \sum_{k=m+1}^n \sigma_{k,dir}~,
\end{equation}
where we have defined
\begin{equation}
\sigma_{k,dir}\equiv -\tau_k L_{kk}X_k \sum_{i=1}^{m}\sum_{j=1}^{m} X_i X_j ~.
\end{equation}
We see that $\sigma_{slow}$ is affected by thermodynamic induction.  Induction also affects the fast variables:
\begin{equation}
\sigma_{fast}= \sum_{k=m+1}^n  L_{kk}X_k^2 + \sum_{k=m+1}^n \sigma_{k,ind}   ~, \label{sigfast2}
\end{equation}
where we have defined the induced rate of entropy production for variable $a_k$ as
\begin{equation}
\sigma_{k,ind} \equiv \tau_k^* L_{kk}X_k \sum_{i=1}^{m}\sum_{j=1}^{m} X_i X_j = -r_k\sigma_{k,dir} ~. \label{sigmakind}
\end{equation}

\subsection{Physical argument for thermodynamic induction}

Below, we will make an argument that the induction term contributes in a significant way to the stability of stationary states. As well, we provide here a physical argument, as thermodynamic induction may at first seem a strange concept;  If two subsystems A and B  are completely decoupled then subsystem B could never be shifted away from equilibrium simply due to a flux caused by subsystem A being out of equilibrium.  In VKC systems the coupling is indirect, and yet we find that a shift in subsystem B can indeed occur because the coupling enters into the expression for $\sigma_T$.
%  It is difficult perhaps to imagine how a given thermodynamic subsystem can be influenced simply because the kinetic coefficient of another thermodynamic subsystem happens to depend on the state of the given system.  
The induction effect is compatible with proper implementation of the all-important factor $\exp(\Delta S_T/k_B)$.  We cite the example of adsorption where it is well known that the chemical potential of an adsorbed species is not the same as the binding energy.  The difference lies in entropic factors, since the adsorbates would have more entropy in the gas phase.  Thus we see entropic-based factors similar to $\exp(\Delta S_T/k_B)$ in expressions for the desorption rate (flux)~\cite{Zangwill}.  In chemical reactions the $\Delta S_T$ value for a reaction plays an important role in deciding the yield.  Reactants can be induced to participate in a reaction to a higher extent if the total system entropy is increased, even if the mechanical variables directly associated with that given reactant are not affected by the entropy increase, but rather the increase in total entropy is distributed among the product states and/or the environment.  For example when two reactant molecules undergo an exothermic reaction at a surface, the excess heat of reaction could wind up in the substrate.  The ensemble of reactants is (statistically) influenced into participating even though it is another system (substrate) that benefits by having its entropy increased.  
% One doesn't get to decide that the $\Delta S$ gets associated with a given subsystem; it's the total total entropy change for the entire system over the entire duration of the process. 

\section{Quasistationary states} \label{statstates}

The concept of stationary states, as described in Refs.~\cite{Prigogine,degroot} in the linear regime, is still applicable in this nonlinear context.  For this discussion, we feel it necessary to make a subtle distinction between (completely) stationary states and quasistationary states.  For quasistationary conditions, the variables $a_i$ evolve slowly in time.  If we formally take the limit where all of the slow timescales $\tau_i$ go to infinity, then we have the conditions for stationary states to exist.  Both types of states likely fall under what Prigogine intended his definition of stationary states to refer to.  Here we feel it is necessary to carefully distinguish between the two types.  For a physical example, an electrical circuit involving a capacitor slowly draining charge (with large RC time constant) corresponds to quasistationary conditions.  Replacing the capacitor with a power supply will hold the applied bias indefinitely over time and would create stationary conditions.   

Following Ref.~\cite{degroot} and defining the fluxes $J_p\equiv \bar{\dot{a}}_p$, then Eq.~(\ref{Neqs}) becomes:
\begin{equation}
J_p=\sum_{q=1}^n (L_{pq}+N_{pq})X_q~~.     \label{Jeqs} 
\end{equation}
We consider the case where $t=0$ corresponds to a system state where all of the slow variables are significantly far away from equilibrium (perhaps caused by a large fluctuation), and all of the fast variables have equilibrium values $(\{X_k=0\})$.  For $t>0$ we wait until all fast variables have had enough time to respond and for all transient response on fast timescales $\tau_k$ to disappear.  Fluxes for the fast variables will also essentially disappear.  Subsequently, these variables will evolve slowly, as quasistationary states, while tracking the slow variables.  In the limit where all slow timescales $\tau_i$ go to infinity, then the fast variables are truly stationary.  Thus we define quasistationary states by the conditions: $J_k=0$, for $k>m$.  Explicitly: 
\begin{equation}
J_k= L_{kk}X_k+ \sum_{i=1}^{m} N_{ki}X_i =0   ~~.      
\end{equation}
Solving and using Eq.~(\ref{Nind}) gives:
\begin{equation}
X_k|_{qss}=- \frac{1}{ L_{kk}}\sum_{i=1}^{m} N_{ki}X_i = - \tau^*_{k}\sum_{i=1}^{m} \sum_{j=1}^{m}\gamma_{ij,k}   X_i X_j  ~.  \label{XfastStat}    
\end{equation}
%One can solve for the $n-m$ fast $X_k$ variables by solving a linear algebraic system $LX=v$, where $L$ is the fast (lower right) block of the full $L_{pq}$ matrix, $X$ is the vector formed from the fast $X_k$ variables only, and $v_k=-\sum_{i=1}^m N_{ki}X_i$ (sum over slow variables only).  Since the matrix $L$ must be invertible, the solution always exists as $X=L^{-1}v$. 

When a given fast variable $x_k$ is quasistationary, it is out of equilibrium, and the entropy of the subsystem associated with that variable is lowered from the equilibrium value by a net amount $\Delta S_k$ given by
%Right after $t=0$, the dissipation rate among fast variables is zero, while the induced rate of entropy production is negative, so $\sigma_{fast}$ becomes negative.  After, it reaches a minimum value before rising and becoming (near) zero when all the quasistationary states for fast variables are achieved.  Over this (fast) time period the net entropy change for each fast variables is given by:
%After a fast variable $x_k$ has achived quasistationary status
\begin{equation}
\Delta S_{k}= \frac{1}{2}X_k a_k =- \frac{1}{2}g_{kk}^{-1}X_k^2 = -\frac{1}{2}g_{kk}^{-1} \left[\tau^*_{k}\sum_{i=1}^{m} \sum_{j=1}^{m}\gamma_{ij,k}   X_i X_j \right]^2 <0 ~. \label{kquartic}
\end{equation}
%The net entropy change for all the fast variables is
%\begin{equation}
%\Delta S_{fast}=\sum_{k=m+1}^n \Delta S_k =-\sum_{k=m+1}^n g_{kk}^{-1}X_k^2 ~.
%\end{equation}
This expression (not to be confused with $\Delta S_T$ from Eq.~(\ref{XjDeltaAj})) is the entropy change for the fast variable considered as an isolated system, so by the second law of thermodynamics this must be a negative definite quadratic form.  
%Explicitly then, in terms of the slow forces, using Eq.~(\ref{XfastStat}) for the quasistationary state we have
%\begin{equation}
%\Delta S_{fast}= -\sum_{k=m+1}^n g_{kk}^{-1}X_{k,qss}^2 =-\sum_{k=m+1}^n g_{kk}^{-1} \left[\tau^*_{k}\sum_{i=1}^{m} \sum_{j=1}^{m}\gamma_{ij,k}   X_i X_j \right]^2 <0  \label{quartic}
%\end{equation}
The fast subsystem can exist in a condition of lower-than-equilibrium entropy for sustained periods of time, as long as at least one of the relevant slow subsystems remains out of equilibrium.  This sustained state would be impossible in thermodynamic equilibrium.  Indeed, it is possible that $-\Delta S_{k}$ could be large enough, so that while in equilibrium, such states would be sampled only for extremely short periods of time during very rare, extreme, fluctuations, i.e., events that are often considered to be thermodynamically inaccessible.  This quartic form in the slow variables $\{X_i\}$ provides a good measure for how far the fast subsystem can be pushed away from equilibrium under sustained conditions. 
% may be considered as a type of \textit{ differential free entropy} since the fast subsystem can borrow or soak up an amount $-\Delta S_{fast}$ during the process where it adjusts to allow the slow subsystem to produce entropy at the higher rate. 
Furthermore, we point out that the expression in Eq.~(\ref{kquartic}) is a differential quantity.

The actual subsystem in question could consist of two distinct spatial regions with an imbalance in the relevant fast variables.  For example, the imbalance could be in thermal energy between two regions of space.  If we focus on just one fast variable $x_k$, then the transfer will lead to one region (for example the one losing thermal energy) having a decrease in entropy $|\delta S_k|$.
% that is much larger in magnitude than $\Delta S_k$.  
In particular, using Eq.~(\ref{XfastStat}) for the quasistationary state we obtain:
\begin{equation}
\delta S_{k}=\frac{\partial S}{\partial a_k}a_k|_{qss}= -\frac{\partial S}{\partial a_k}g_{kk}^{-1}X_k|_{qss}=-\frac{\partial S}{\partial a_k}g_{kk}^{-1}\tau^*_{k}\sum_{i=1}^{m} \sum_{j=1}^{m}\gamma_{ij,k}X_i X_j~.    \label{deltaSk}
\end{equation}
This quantity can be positive or negative, depending on the signs of the $\gamma_{ij,k}$ coefficients.  We use the phrase \textit{free entropies} for these $\delta S_{k}$ terms.  In the thermal energy transfer example, $T\delta S_k$ would be the actual amount of heat induced to transfer between the two regions of the fast subsystem.  Note that no heat needs to be transferred between fast and slow subsystems.  If the magnitude of this quantity becomes significant, relative to the equilibrium entropy of the region in question, then the quasistationary states could be very interesting, and perhaps very difficult to attain otherwise.  For example, these states could involve some level of self-organization~\cite{Nicolis1977}.  

We feel the $\delta S_{k}$ is a better measure than the differential entropy expression of Eq.~(\ref{kquartic}), in assessing the potential for producing such interesting effects. Multiplying $\delta S_k$ by the local temperature gives a new type of free energy that provides one with an energy budget available for overcoming barriers keeping this subsystem from making transitions to these interesting quasistationary states.

The reduction in entropy caused by shifting the fast variables into quasistationary states suggests that the entropy production associated with the slow variables may be increased.  We will indeed find this to be the case, but beforehand, an investigation of the stability of quasistationary states is warranted.

\subsection{Stability of quasistationary states}

The key point for the linear stability analysis is realizing, from Eq.~(\ref{ainfexplicit}), that no fast variables $X_k$ are present in the induction terms.  Thus if a small change in a single fast variable $X_l$ is made i.e. $\delta X_l$, then the change in the flux $J_l$ is the same as it is in the linear case.  Since $J_l=0$ in the quasistationary state, we have $J_l=L_{ll}\delta X_l$.  The argument for stability follows exactly as it does for the linear case (see Ref.~\cite{degroot}) i.e. the positivity of $L_{ll}$, as well as the positive-definite character of $g_{pq}$, guarantees that if $a_l$ is pushed away from the stationary state, then the sign of $J_l=\dot{a}_l$ will be such as to return the variable back towards the quasistationary state.  Thus, it appears that the fast states are all stable when quasistationary.

\subsubsection{Nonlinear stability analysis}

This argument can fail when the linear kinetic coefficients are very small.  In such a case it is possible that the nonlinear terms dominate and decide the issue of stability.  To illustrate we consider the case with one slow and one fast variable i.e. $n=2$, $m=1$.  The dynamical equations~(\ref{Neqs}) become
\begin{equation}
{\dot{a}}_1 = L_{11}X_1 + (c X_1) X_2  ~,
\end{equation}
\begin{equation}
{\dot{a}}_2=  - (r c X_1) X_1 + L_{22}X_2  ~. 
\end{equation}
Using Eq.~(\ref{agX}):
\begin{equation}
\dot{X}_1 = -\frac{L_{11}}{g_{11}}X_1 - (\alpha X_1) X_2  ~,
\end{equation}
\begin{equation}
\dot{X}_2=  + r(\alpha X_1) X_1 - \frac{L_{22}}{g_{22}} X_2  ~. 
\end{equation}
%Now suppose $\tau_2$ is large and $\tau_1>>\tau_2$ is very large, and $\alpha$ ? % The parameter $r=1$ according to our results.  
%Here we shall find it instructive to allow r to vary.
The quasistationary state for the fast variable corresponds to
\begin{equation}
 r\tau_2 \alpha X_1^2 = X_{2_{ss}}  ~. 
\end{equation}

% Technically the stationary state does change in time at the timescale $\tau_1=\frac{g_{11}}{L_{11}}$. 
In order to facilitate the analysis we assume the slow system is almost static and we ignore the time variation in $X_1$.  We then ask what happens if we make a small change $\delta X_2$ in $X_2$ away from $X_{2_{ss}}$  i.e. $X_2=X_{2_{ss}}+\delta X_2$.  Ignoring the linear (stabilizing terms) and focusing on the nonlinear terms, then~\footnote{We also make the special condition that $M_{11}$ is very small (but still positive) in the stationary state, i.e., $\dot{X}_1\approx 0$ in the stationary state.}:
\begin{equation}
\dot{X}_1 \approx - (\alpha X_1) \delta X_2  ~.
\end{equation}
Over a short time $t$:
\begin{equation}
X_1(t)-X_1(0) \approx - (\alpha X_1) \delta X_2 t ~,
\end{equation}
\begin{equation}
X_1^2(t) \approx X_1^2(0) - 2\alpha X_1^2(0)  \delta X_2 t ~,
\end{equation}
\begin{equation}
\dot{X}_2=  + r\alpha (X_1^2(0) - 2\alpha X_1^2(0)  \delta X_2 t) -\frac{L_{22}}{g_{22}} (X_{2_{ss}}+\delta X_2)  =  -2 r\alpha^2 X_1^2(0)  \delta X_2 t -\frac{L_{22}}{g_{22}}\delta X_2   ~. 
\end{equation}

\begin{figure}[ht]%
\includegraphics[width=0.8\columnwidth]{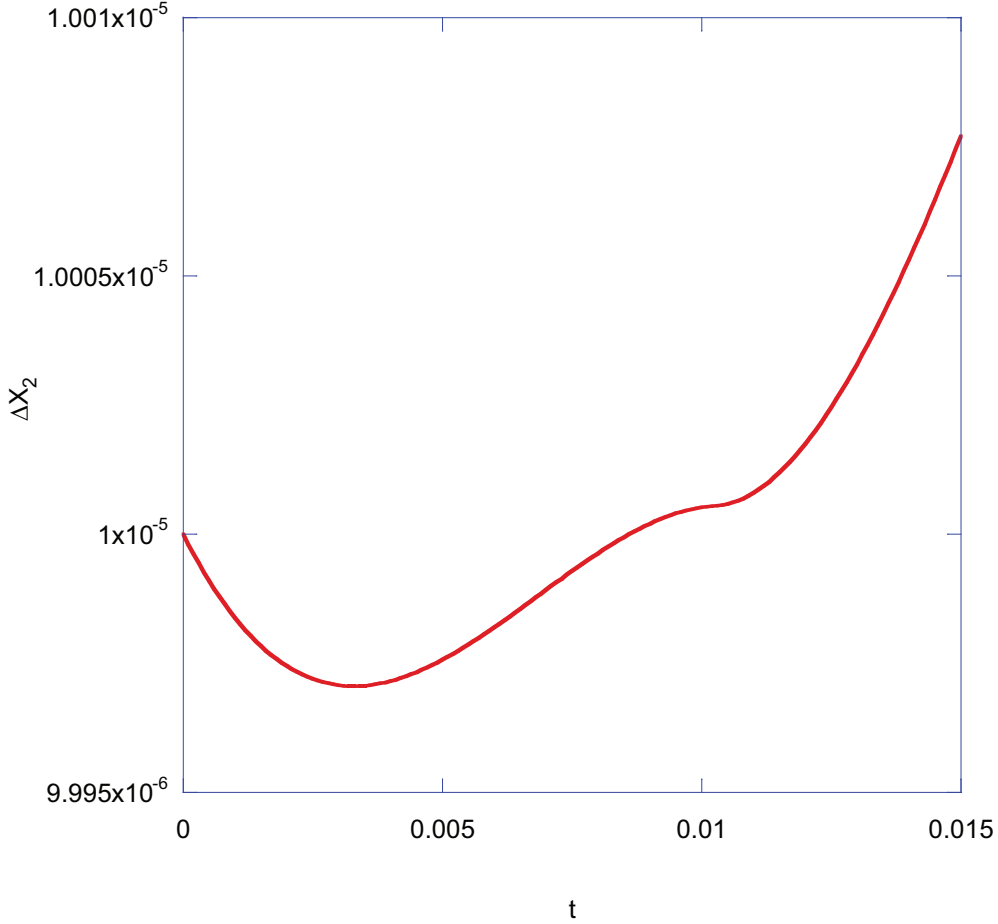}%
\caption{Plot of $X_2(t)-X_{2_{ss}}$ with parameter $r$ having the incorrect sign $(r=-1)$.  This solution displays an unstable quasistationary state. }%
\label{fig:unstable}%
\end{figure}

For positive $r$ we see that the nonlinear term reinforces the stabilizing effect of the linear term.  However, for negative $r$ the quasistationary state can become unstable. Technically, for small enough values of $t$ the stability is there, but realistically this time may have to be extremely small.  For reasonable and relevant timescales the nonlinear term could win out and cause instability.  It's not very difficult to set up an  example problem with explicit numerical solution illustrating instability.  
%For example, $X_(0)=1$, $g_{11}=g_{22}=1$, $\tau_1=10^6$, $\tau_2=100$, $\alpha=0.5$, $r=-1$, $\delta X_2=10^{-4}$ gives the solution shown in Fig.~\ref{fig:unstable} .  The stationary state occurs at $X_2=-0.2$.
For example, $X_1(0)=1$, $L_{11}=0.051$ $g_{11}=g_{22}=1$, $L_{22}=0.2$, $\alpha=0.1$, $r=-1$, $\delta X_2=10^{-5}$ gives the solution for $\Delta X_2\equiv X_2(t)-X_{2_{ss}}$ shown in Fig.~\ref{fig:unstable} .  The quasistationary state occurs at $X_2=-0.5$.  We see that for very small times on the order of $t=0.001$ s the response appears to be a stabilizing return back to zero i.e. $X_2=-0.5$.  But the system never returns to the quasistationary state.  At around $t=0.003$ s, $X_2(t)$ turns around and then responds in a way consistent with instability.
We conclude that if there is to be an induction term then we have indeed obtained the correct sign i.e. $r>0$.  The $r=-1$ option clearly causes problems with stability.  Even smaller, negative values of $r$ will also cause instabilities.  In a certain sense, then $r=0$ can be thought of as a situation neither stable or unstable i.e. neutral.  We feel that this further justifies the case for the induction effect: it may be true in general that in systems approaching equilibrium and containing kinetic coefficients that depend on other thermodynamic variables, the induction term is necessary for providing stable approaches to equilibrium via quasistationary states.

\subsection{Fluxes and entropy production}

%With no loss of generality we may consider a change of basis to fast (eigenvector) variables such that $L$ is diagonal.  This will aid in uncovering an important inequality pertaining to rates of entropy production, as follows.  
%the forces for fast variables are given by:
%\begin{equation}
%X_k =-\frac{1}{L_{kk}}\sum_{i=1}^m N_{ki}X_i   ~.  \label{XfastStat}    
%\end{equation}
If we focus on one fast variable $a_k$ while in a quasistationary state, we see that $\sigma_k=0$ and yet the variable is not in it's equilibrium state.  This means that $L_{kk}X_k^2>0$.  We refer to such a term as a {\it dissipative rate of entropy production}, or {\it dissipation rate}, for short.  The term is apt since its physical origin is strict relaxation.
% the dissipation rate via the linear $L_{kk}$ kinetic coefficient is non-zero.  This dissipation creates entropy at the rate $L_{kk}X_k^2>0$. 
While in the stationary state, the entropic induction continually pushes the variable in a direction opposite to dissipation, as far as entropy production goes. 
% We shall use the term \textit{organization} to describe this direction opposite to dissipation.   
%In the stationary state there is a balance between dissipation and organization, that gives zero net entropy production. 
%Using Eq.~(\ref{sigmakind}), we define then, the \textit{induced rate of entropy production} for variable $a_k$ as
%\begin{equation}
%\sigma_{ind,k}=  \sum_{i=1}^{m}\sum_{j=1}^{m} \gamma_{ji,k} L_{kk}\tau_k^* X_i X_j X_k ~. \label{orgrate}
%\end{equation}
Referring to Eq.~(\ref{sigmakind}) we note that, unlike the dissipation rate, the induced rate of entropy production for variable $a_k$ can be positive or negative, depending on the precise state and also on the signs of the $\gamma_{ji,k}$ coefficients.  However, in the quasistationary state there is a balance between induction and dissipation: $\sigma_{k,ind,qss}=-L_{kk}X_k^2$. The quasistationary induced rate of entropy production is always negative.  Given the generality of this thermodynamic approach, there will likely be many physical interpretations for such a negative value.  Self-organization may well be one of them.
%This is a negentropy rate.  While in the stationary state, the rate of organization is positive definite and $O_{k,ss}=L_{kk}X_k^2$.

\textbf{Theorem 2:}

If $\sigma_T |_{\{X_k=0\}}$  is the total entropy production with all fast variables at their equilibrium values, and $\sigma_T |_{\{J_k=0\}}$  is the total entropy production with all fast variables quasistationary then    
\begin{equation}
\sigma_T |_{\{J_k=0\}} -\sigma_T |_{\{X_k=0\}} = \sigma_{extra} \geq 0~,      
\end{equation}
where
\begin{equation}
\sigma_{extra}=\sum_{k=m+1}^n g_{kk}\tau^*_k\left(\sum_{i=1}^{m} N_{ik}X_i\right)^2 ~.    \label{sigmaex}
\end{equation}

To prove this theorem we note that quasistationary states do actually evolve slowly over time while the slow variables gradually relax towards equilibrium.  Substituting Eq.~(\ref{XfastStat}) into Eq.~(\ref{Jeqs}), and with all the fast variables in quasistationary states, the fluxes for the slow variables are given by:
\begin{equation}
J_i =\sum_{j=1}^m L_{ij}X_j+ \sum_{k=m+1}^n N_{ik}X_k =\sum_{j=1}^m L_{ij}X_j - \sum_{k=m+1}^n N_{ik}\frac{1}{L_{kk}}\sum_{j=1}^m N_{kj}X_j ~~.      
\end{equation}
With all fast variables quasistationary, $\sigma_{fast}=0$.  Using Eqs.~(\ref{gLtauOne}),~(\ref{relateIndDir}),~(\ref{sigmakind}),~(\ref{XfastStat}) we find
\begin{equation}
\sigma_T |_{\{J_k=0\}}=\sigma_{slow} =\sum_{i=1}^m\sum_{j=1}^m L_{ij}X_i X_j+ \sigma_{extra}  ~,   \label{sigTstat}   
\end{equation}
 i.e., the desired result.
The extra entropy production rate $\sigma_{extra}$ is positive definite in the quasistationary state.  

The physical meaning for this result is that the entire system produces entropy faster when the fast variables are allowed to relax by moving away from equilibrium values and achieving quasistationary status.  In the quasistationary state we may think of the quantity $\sigma_{extra}$ as the increase in $\sigma_{slow}$ on top of the linear contribution.  This result constitutes a nonequilibrium version of Le Chatelier's principle.  In the traditional Le Chatelier's principle, which is an equilibrium thermodynamics principle, when a given thermodynamic variable is pushed away from equilibrium, other thermodynamic variables (that are coupled by the $g$ matrix) relax to new equilibrium values, so that the total entropy is again maximized, and the new relaxed entropy is always greater than the unrelaxed entropy~\cite{LandauL}.  Here, when we push a slow variable away from equilibrium, fast variables (that are coupled by $\gamma_{ij,k}$) will temporarily relax away from equilibrium to quasistationary states, and the new relaxed rate of entropy production is always greater than the unrelaxed rate.  Note that our distinction between slow and fast states is essential in arriving at this new principle.  We point out that the stationary state limit, where all variables become frozen in time, is not thermodynamic equilibrium, since the slow fluxes $J_i$ are not zero.  This is worth pointing out since one might mistakenly conclude thermodynamic equilibrium if one focuses only on the fast variables.

%The effect i.e. $\sigma_{extra}$ disappears when the entire system reaches equilibrium, so the result is distinct from the traditional Le Chatelier's principle.
%, which is an equilibrium thermodynamics principle.  These equilibrium effects are built into the g matrix .  Theorem 2 adds on to the equilibrium result, a shift away from equilibrium in the fast variables, caused by the slow variables.  The amount of the shift depends quadratically on the slow variables.  Instead of maximizing ST, we maximize Phi.

%, and 2) for the fast variables $\sigma_{ex}$ exactly balances the entropy production from the linear dissipation i.e. $\sum_{k=m+1}^n\sum_{l=m+1}^n L_{kl}X_k X_l =\sigma_{ex}$.  

Since we have shown that allowing the fast variables to relax to quasistationary states leads to increased overall entropy production, we are led to formulate a variational principle, which maximizes entropy production in a certain sense. 
%  we may ask if the total rate of entropy production can be maximized, anticipating a type of variational principle for maximum entropy production.
We can use this variational principle to better understand why fast variables would shift away from their equilibrium values $a_k=0$.

%The positive quantity $-\delta S_k$ is termed the \textit{free negentropy}
  %We now ask an important question:  For what reason would such subsystems spontaneously move away from equilibrium?
%
%In order to help answer this question, we note that the entire system reaches equilibrium sooner when fast states are in the stationary state configurations.  The time needed for the fast variables to become stationary is negligible as compared to the relaxation time for slow variables, and these slow variables determine the time required to achieve complete equilibrium.    

\section{Variational principles for entropy production} \label{sec:maxent}

%Previous attempts to formulate a maximum entropy principle have not succeeded beause they try to fix all the variables Xi.  
We follow the approach taken by Prigogine, for linear systems, where the choice is made to hold constant some, but not all, variables while leaving the rest free to vary~\cite{degroot}.  In our case the slow variables play the role of Prigogine's fixed variables, as viewed by the fast variables, which play the role of the free variables.  For the linear system one minimizes the total rate of entropy production~\cite{degroot}.  For the nonlinear case considered here there are some differences.  Using $\sigma_{slow}$ as a starting point, we introduce the \textit{free entropy production} (rate) as
\begin{equation}
\Phi \equiv  \sum_{i=1}^{m}\sigma_i -\sum_{k=m+1}^n r_k^{-1} \sigma_k = \sum_{i=1}^{m}J_i X_i -\sum_{k=m+1}^n r_k^{-1}J_k X_k = \sigma_{slow} -\sum_{k=m+1}^n r_k^{-1}J_k X_k~. \label{PhiDef}
\end{equation} 
The free entropy production clearly differs, in general, from the total entropy production.  We formulate the following theorem.

\textbf{Theorem 3: (principle of maximum free entropy production)}

When the free entropy production, $\Phi$, is maximized, the fluxes for all the fast variables vanish, i.e., $J_k=0$ for $k>m$.

To prove this theorem we first substitute into Eq.~(\ref{PhiDef}) for the fluxes, using Eq.~(\ref{Jeqs}) to give 
%In terms of linear contributions and $\sigma_{ex}$, using Eq.~(\ref{sigmaEx}):
%\begin{equation}
%\sigma_F = \sum_{k=1}^{m}\sum_{l=1}^{m}L_{kl} X_k X_l + \sum_{k=1}^{m}\sum_{i=m+1}^n N_{ki} X_k X_i -\sum_{i=m+1}^n\sum_{j=m+1}^n L_{ij} X_i X_j - \sum_{k=1}^{m}\sum_{i=m+1}^n N_{ik} X_i X_k
%\end{equation}
%\begin{equation}
%\sigma_F = \sum_{i=1}^{m}\sum_{j=1}^{m}L_{ij} X_i X_j -\sum_{k=m+1}^n\sum_{l=m+1}^n L_{kl} X_k X_l  +2 \sigma_{ex} ~.
%\end{equation}
%\begin{equation}
%\Phi = \sum_{i=1}^{m}\sum_{j=1}^{m}L_{ij} X_i X_j +\sum_{i=1}^{m}\sum_{k=m+1}^n N_{ik} X_i X_k -\sum_{k=m+1}^n  r_k^{-1} L_{kk} X_k^2  - \sum_{i=1}^{m}\sum_{k=m+1}^n r_k^{-1} N_{ki} X_i X_k ~.
%\end{equation}
%\begin{equation}
%\Phi = \sum_{i=1}^{m}\sum_{j=1}^{m}L_{ij} X_i X_j -\sum_{i=1}^{m}\sum_{k=m+1}^n r_k^{-1} N_{ki} X_i X_k -\sum_{k=m+1}^n  r_k^{-1} L_{kk} X_k^2  - \sum_{i=1}^{m}\sum_{k=m+1}^n r_k^{-1} N_{ki} X_i X_k ~.
%\end{equation}
\begin{equation}
\Phi = \sum_{i=1}^{m}\sum_{j=1}^{m}L_{ij} X_i X_j -2 \sum_{i=1}^{m}\sum_{k=m+1}^n r_k^{-1} N_{ki} X_i X_k -\sum_{k=m+1}^n r_k^{-1} L_{kk} X_k^2   ~.
\end{equation}
%Explicitly written in terms of the variables $X_k$ then:
%\begin{eqnarray}
%\sigma_F &=& \sum_{i=1}^{m}\sum_{j=1}^{m}L_{ij} X_i X_j -\sum_{i=m+1}^n L_{ii} X_i^2 -\sum_{i=m+1}^n\sum_{j\neq i} L_{ij} X_i X_j \nonumber \\
%&-& 2 \sum_{k=1}^{m}\sum_{i=m+1}^n X_k X_i\sum_{l=1}^m \sum_{j=m+1}^{n}\gamma_{kl,j} g_{ji}^{-1}X_l~.
%\end{eqnarray}
Maximizing with respect to each fast variable, and recalling that the $N_{ik}$ coefficients do not depend on any fast variables $X_k$, results in $n-m$ conditions:
\begin{eqnarray}
\frac{\partial \Phi}{\partial X_k} =  -2 r_k^{-1} \sum_{i=1}^{m} N_{ik} X_i -2 r_k^{-1} L_{kk} X_k = -2 r_k^{-1} J_k =0~,
\end{eqnarray}
where comparison to Eq.~(\ref{ainfexplicit}) has been made.  Thus, we have proven that stationary states maximize $\Phi$.  When all of the $n-m$ fast states are stationary (definition of completely stationary) then $\Phi=\sigma_{slow}=\sigma_T$.  As shown in Sec.~\ref{statstates}, this total rate of entropy production with all $J_k=0$ is larger than if all fast variables $X_k$ were zero. 

\textbf{Corollary 1:}

One can easily show, using the method of Lagrange multipliers, that the quantity $\sigma_{slow}$ is maximized when the fast variables $X_k$ take their quasistationary values, if we also add the $n-m$ constraints: $\sigma_{k}=0$ for $k>m$, i.e. for all fast states.  The Lagrange multipliers are identified as $r_k^{-1}$.  

\textbf{Corollary 2: (principle of maximum entropy production)}

Also, the  total entropy production $\sigma_{T}$ is maximized when the fast variables $X_k$ take their quasistationary values, with the same $n-m$ constraints: $\sigma_{k}=0$ for $k>m$.  In this case, the Lagrange multipliers are $1+r_k^{-1}$.  

The quasistationary states are very important since they maximize the total entropy production, as long as we understand the constraints and that only fast variables are involved in the maximization procedure. In this sense, we may now refer to quasistationary states also as states of maximum entropy production.

Thus far we have formulated maximum entropy production principles in three ways.  A fourth formulation is obtained as follows. If we take the limiting procedure where slow state variables are actually fixed then we arrive at the following principle: when a system described by n variables is held in a state with fixed $X_1,~X_2,...,X_m$ (with $m<n$) and maximum free entropy production $\Phi$, then the fluxes $J_k$ with $m<k\leq n$ vanish. 

%One can easily show that the same result is obtained when maximizing $\sigma_{slow}$ with the $n-m$ constraints: $\sigma_{k}=0$ for $k>m$, i.e. for all fast states.  The Lagrange multipliers are $r_k^{-1}$.  Equivalently, the same result is obtained when maximizing the total entropy production, $\sigma_{T}$, with the same $n-m$ constraints: $\sigma_{k}=0$ for $k>m$.  In this case, the Lagrange multipliers are $1+r_k^{-1}$.  

When all of the $n-m$ fast states are stationary, the maximal value of the free entropy rate is given by Eq.~(\ref{sigTstat}):   
\begin{equation}
\Phi_{max}=\sigma_{slow}|_{\{J_k=0\}}=\sigma_T|_{\{J_k=0\}} =\sum_{i=1}^m\sum_{j=1}^m L_{ij}X_i X_j+ \sigma_{extra}  ~.   \label{SigmaMax}   
\end{equation}
Physically, one thinks now of more than just fast variables relaxing to nonequilibrium values; The fast variables adjust themselves so that the whole system gets to equilibrium faster.  In fact, in the quasistationary states the whole system approaches equilibrium as fast as possible, given some important restrictions.  The extent of the adjustment of the fast variables must have limitations.  Maximizing $\sigma_{slow}$ or $\sigma_T$ without any constraints on the forces $X_k$ would give an unphysical runaway result.  The fast variables relax until they become quasistationary and an important dynamical balance is achieved.  This balance is the reason for the minus signs in front of the fast variable $\sigma_k$ terms in Eq.~(\ref{PhiDef}).

While the fast variables adjust themselves so that the whole system gets to equilibrium faster, they may spend considerable time in states with lower entropy than their equilibrium states ($a_k=0$) would have, i.e. $\Delta S_k<0$. The complex states and interesting structures suggested above could be created as the fast variables sample their phase space and seek configurations that maximize the rate of entropy production of the slow system (with the constraints $\sigma_{k}=0$ for $k>m$).
% could discuss PEAPP here, system does it's best to obey PEAPP as it approaches but before reaching equil. some subsystems may have unequal weighting of microstates.
This effect should be enhanced if the slow system is large while the fast system is small and has a gating, or bottlenecking, property of strongly controlling the pertinent kinetic coefficients of the large system.

\subsection{Stationary states}

If we take the stationary limit, $\tau_i \rightarrow\infty$, where the slow variables are held constant, then the coefficients $N_{pq}$ become constants.  The slow variables are essentially projected out of the problem, acting as passive reservoirs.  We note that at least one slow variable is required to play the essential role of dynamical reservoir.  We note that Theorems 1-3 and corollaries still apply in the stationary limit.    The thermodynamic induction effect persists as constant terms in the remaining $n-m$ dynamical equations for the fast variables. These terms serve to drive the fast variables away from equilibrium, and towards the stationary states.  The dynamical equations for the fast variables become linear.  It is remarkable that a problem that begins unavoidably as nonlinear, becomes linear in this particular limit.

\section{Case of two variables} \label{sec:2var}

To help illustrate these concepts with an example, we consider the simplest system possible that exhibits thermodynamic induction, in particular, the case where $n=2$ and $m=1$, i.e., one slow variable acting as the dynamical reservoir, and one fast variable.  This is an important case to consider since it likely suffices to cover many applications of thermodynamic induction.  This case illustrates the essential features of the reservoir variable interacting with a variable that has its dynamics coupled to the reservoir.  In the simplest case, these two variables would be completely uncoupled except that the kinetic coefficient for the slow variable $(i=1)$ happens to depend on $a_2$.  This means the two variables are uncoupled up to linear order i.e. $g_{12}=g_{21}=0$ and $L_{12}=L_{21}=0$.  This must be the case since if, for example, $g_{12}$ was nonzero, then we could not have one very slow timescale $\tau_1$ and one very fast timescale $\tau_2$.  We can also see this as a consequence of Assumption 2.    The coupling at the nonlinear level will be described by the coefficient $\gamma_{11,2}$ as prescribed in assumption 1.  We note that Assumption 2 guarantees that $\gamma_{12,1}=\gamma_{21,1}0=\gamma_{12,2}=\gamma_{21,2}=0$, while Assumption 4 sets $\gamma_{11,1}=0$.  Lastly, $\gamma_{22,1}=\gamma_{22,2}=0$ by Assumption 3.  Thus, under our set of assumptions, only $\gamma_{11,2}$ can be nonzero.   For the slow variable, Eq.~(\ref{Ndir}) gives $N_{12}=-\gamma_{11,2} g_{22}^{-1}X_1$, which creates the coupling between the two variables, and Eqs.~(\ref{Neqs}) become
\begin{equation}
{\bar{\dot{a}}}_1= L_{11}X_1  -  \gamma_{11,2} g_{22}^{-1}X_1  X_2   \label{2vardyneq1}
\end{equation}
and
\begin{equation}
{\bar{\dot{a}}}_2= L_{22}X_2 +r_2\gamma_{11,2} g_{22}^{-1} X_1^2 ~. \label{2vardyneq2}
\end{equation}
%or
%\begin{equation}
%\bar{\dot{a}}_1= -L_{11}g_{11}a_1  - g_{22} \gamma_{11,2} g_{22}^{-1}g_{11}a_1 a_2 = -L_{11}g_{22}a_2  -\gamma_{11,2} g_{11}a_1 a_2  
%\end{equation}
%Using $r_2=g_{22}L_{22}\tau_k^*$, for the fast variable, Eq.~(\ref{adotXXfast}),  becomes
%\begin{equation}
%\bar{\dot{a}}_2= -L_{22}g_{22}a_2 +r_2\gamma_{11,2} g_{22}^{-1}g_{11}^2 a_1^2 ~.  
%\end{equation}

Thus we verify our claims made in Eqs.~(\ref{claim1}),~(\ref{claim2}), (with $c= -\gamma_{11,2} g_{22}^{-1}$).  The induction term manifests itself as $+r_2\gamma_{11,2} g_{22}^{-1} X_1^2$ and affects the dynamics of $a_2$.  Thermodynamic equilibrium corresponds to $a_1=a_2=0$.  If the slow variable $a_1$ is pushed away from zero (perhaps by a large fluctuation), we see that $a_2$ will be induced to also move away from zero.  After a long time passes, both variables will relax back to equilibrium.  We note that in the linear limit we ignore $\gamma_{11,2}$ and the two timescales are identified as $\tau_1=\tau_{slow}=1/(L_{11}g_{11})$ and  $\tau_2=\tau_{fast}=1/(L_{22}g_{22})\ll\tau_1$.

For the entropy production rates:
\begin{equation}
\sigma_{slow}=\sigma_1= {\bar{\dot{a}}}_1 X_1 = L_{11}X_1^2  -  \gamma_{11,2} g_{22}^{-1}X_1^2  X_2~,
\end{equation}
\begin{equation}
\sigma_{fast}=\sigma_2= {\bar{\dot{a}}}_2 X_2 = L_{22}X_2^2  +  r_2\gamma_{11,2} g_{22}^{-1}X_1^2  X_2~,
\end{equation}
\begin{equation}
\sigma_T=\sigma_{slow}+\sigma_{fast} = L_{11}X_1^2 + L_{22}X_2^2 -(1-r_2)  \gamma_{11,2} g_{22}^{-1}X_1^2  X_2  ~.
\end{equation}
For the free entropy production:
\begin{equation}
\Phi=\sigma_{slow}-r_2^{-1}~\sigma_{fast} = L_{11}X_1^2 - r_2^{-1}  L_{22}X_2^2   - (1+ r_2^{-1} r_2) \gamma_{11,2} g_{22}^{-1}X_1^2  X_2~.
\end{equation}
We verify that
\begin{equation}
\frac{\partial \Phi}{\partial X_2}= - 2r_2^{-1} L_{22}X_2   - 2 \gamma_{11,2} g_{22}^{-1}X_1^2  = -2 r_2^{-1}{\dot{a}}_2~.
\end{equation}
It is clear that maximizing $\Phi$, by varying $X_2$, is equivalent to setting $J_2=0$, i.e., the principle of maximum free entropy production is verified.

\subsection{Quasistationary state}

The one quasistationary state available for this system comes from setting $\bar{\dot{a}}_2=0$.  Thus we maximize the function $\Phi(X_1,X_2)$ by holding $X_1$ constant while varying $X_2$.  In terms of what is actually a slowly varying force $X_1$:
\begin{equation}
X_2 =- \frac{r_2\gamma_{11,2} g_{22}^{-1}}{L_{22}} X_1^2 ~, \label{X2stat}
\end{equation}
\begin{equation}
a_2 = \frac{r_2\gamma_{11,2}}{L_{22}g_{22}^2} X_1^2 ~. 
\end{equation}
By Eq.~(\ref{SigmaMax}) the free entropy rate is
\begin{equation}
\Phi_{max}=\sigma_T |_{\{J_k=0\}} = L_{11}X_1^2 + \sigma_{extra}  ~,     
\end{equation}
where by Eq.~(\ref{sigmaex})
\begin{equation}
\sigma_{extra}=g_{22}\tau_2^*\left( N_{21}X_1\right)^2 = g_{22}\tau_2^*\left( \gamma_{11,2} g_{22}^{-1}X_1^2\right)^2  ~.   
\end{equation}
By Eq.~(\ref{sigmakind}) the induced rate of entropy production is
\begin{equation}
\sigma_{ind,2}= -\tau_2^* r_2\gamma_{11,2}^2 g_{22}^{-1} X_1^4 = -r_2 \sigma_{ex} ~.
\end{equation}
In this state the differential change in entropy for the fast variable is evaluated using Eq.~(\ref{kquartic}):
\begin{equation}
\Delta S_{2}=X_2 a_2 =- \frac{r_2^2\gamma_{11,2}^2}{L_{22}^2 g_{22}^3} X_1^4 =-\tau_2^*\sigma_{extra} ~.
\end{equation}
and we verify that this is negative.

\subsection{Solution of coupled differential equations}

\begin{figure}[ht]%
\includegraphics[width=0.8\columnwidth]{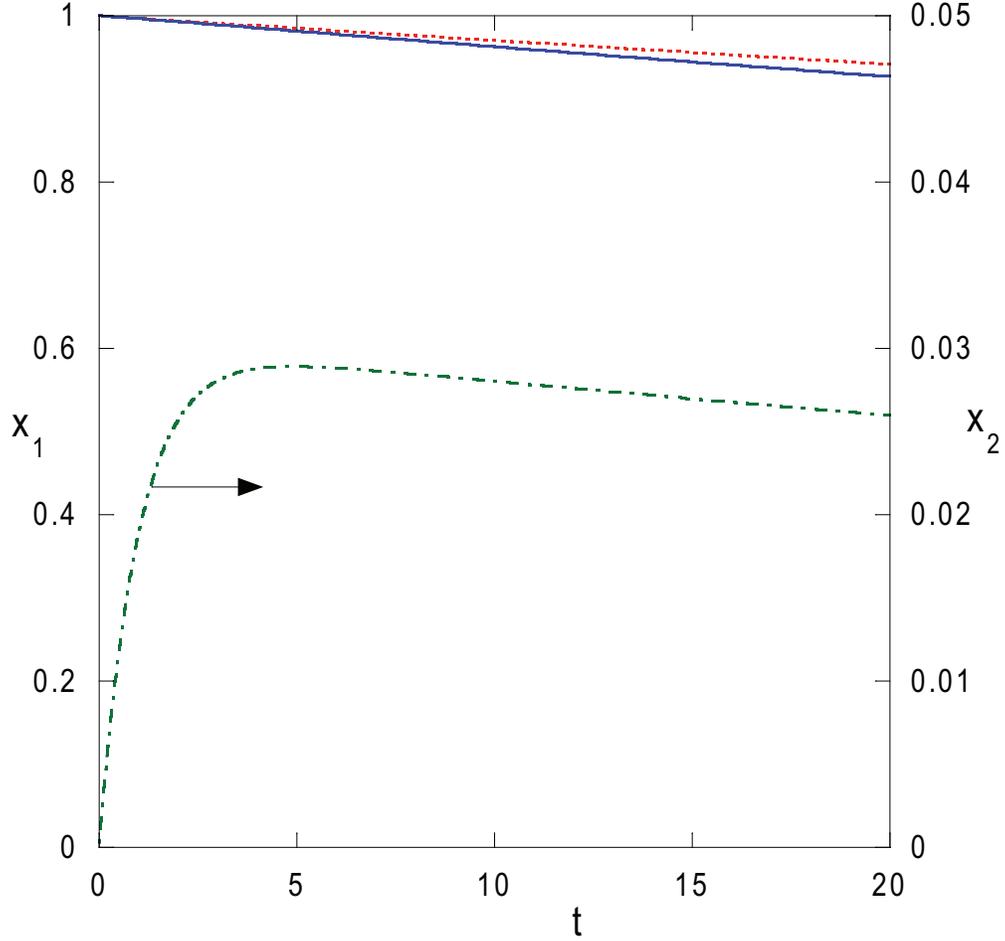}%
\caption{Plots of $X_1(t)$ (solid curve) and $X_2(t)$ (dot-dashed).  Also shown is the solution for $X_1$ with $\gamma_{11,2}$ set to zero (short-dashed). }%
\label{FigDE}%
\end{figure}
\begin{figure}[ht]%
\includegraphics[width=0.8\columnwidth]{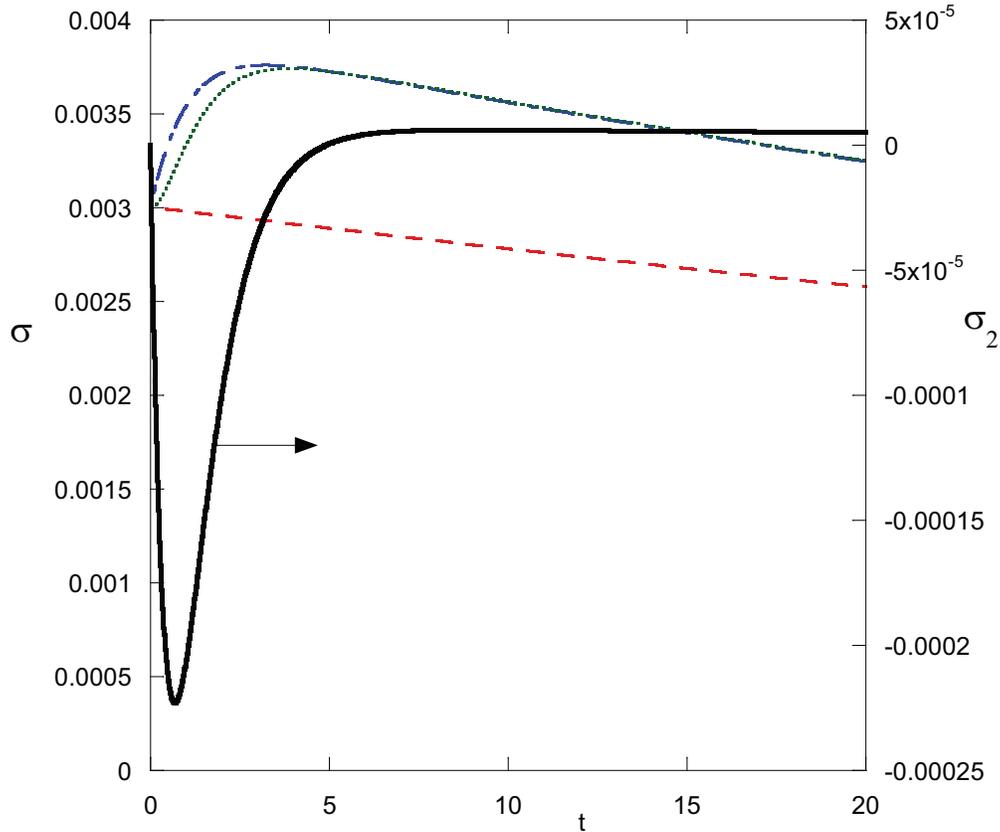}%
\caption{Plots of $\sigma_1(t)$ (dot-dashed curve), $\sigma_2(t)$ (solid) and $\sigma_T(t)$ (short-dashed).  Also shown is $\sigma_T(t)$ (long-dashed) solved with $\gamma_{11,2}$ couplings set to zero.}%
\label{FigSigma}%
\end{figure}

%The $n$ dynamical equations are ordinary homogeneous nonlinear first order differential equations, represented in Eqs.~(\ref{slowdyn},~ \ref{fastdyn}).  In particular, 
For the two variable example discussed here, Eqs.~(\ref{2vardyneq1}),~(\ref{2vardyneq2}) are easily solved numerically.  Here we present an example with the following two dynamical equations, with coupling parameters equal to $\pm 0.03$, ($r=1$):
\begin{eqnarray}
\dot{X}_1 &=& -0.003 X_1 -0.03 X_1 X_2 \nonumber \\
\dot{X}_2 &=& +0.03 X_1^2 - X_2 ~.
\end{eqnarray}
Solutions for given initial conditions $X_1(0)=1$, $X_2(0)=0$, are shown in Fig.~\ref{FigDE} which plots $X_1(t)$ (solid curve) and $X_2(t)$ (dot-dashed).  Also shown (short-dashed), is the solution for $X_1$ with $\gamma_{11,2}$ set to zero.  The coupling increases the rate of approach towards equilibrium for the slow state i.e. the dynamical reservoir.  Variable 2 responds quickly and gets pushed away from equilibrium.  After  $t\approx 5$ the system is in a good approximation to a stationary state.  In Fig.~\ref{FigSigma} we see how the entropy production rates vary.  Without the coupling terms, $\sigma_2$ would be zero and $\sigma_1(t)$ would follow the long dashed curve.  With the couplings, the entropy production for system 2 can be negative for some time.  Of course, the total entropy production never becomes negative.  In fact, $\sigma_T$ (short-dashed) increases to higher levels than for the dashed curve.  This is consistent with the total system approaching equilibrium faster with the coupling terms.

\section{Numerical Simulation}  \label{sec:numer}

\begin{figure}[h]%
\includegraphics[width=\columnwidth]{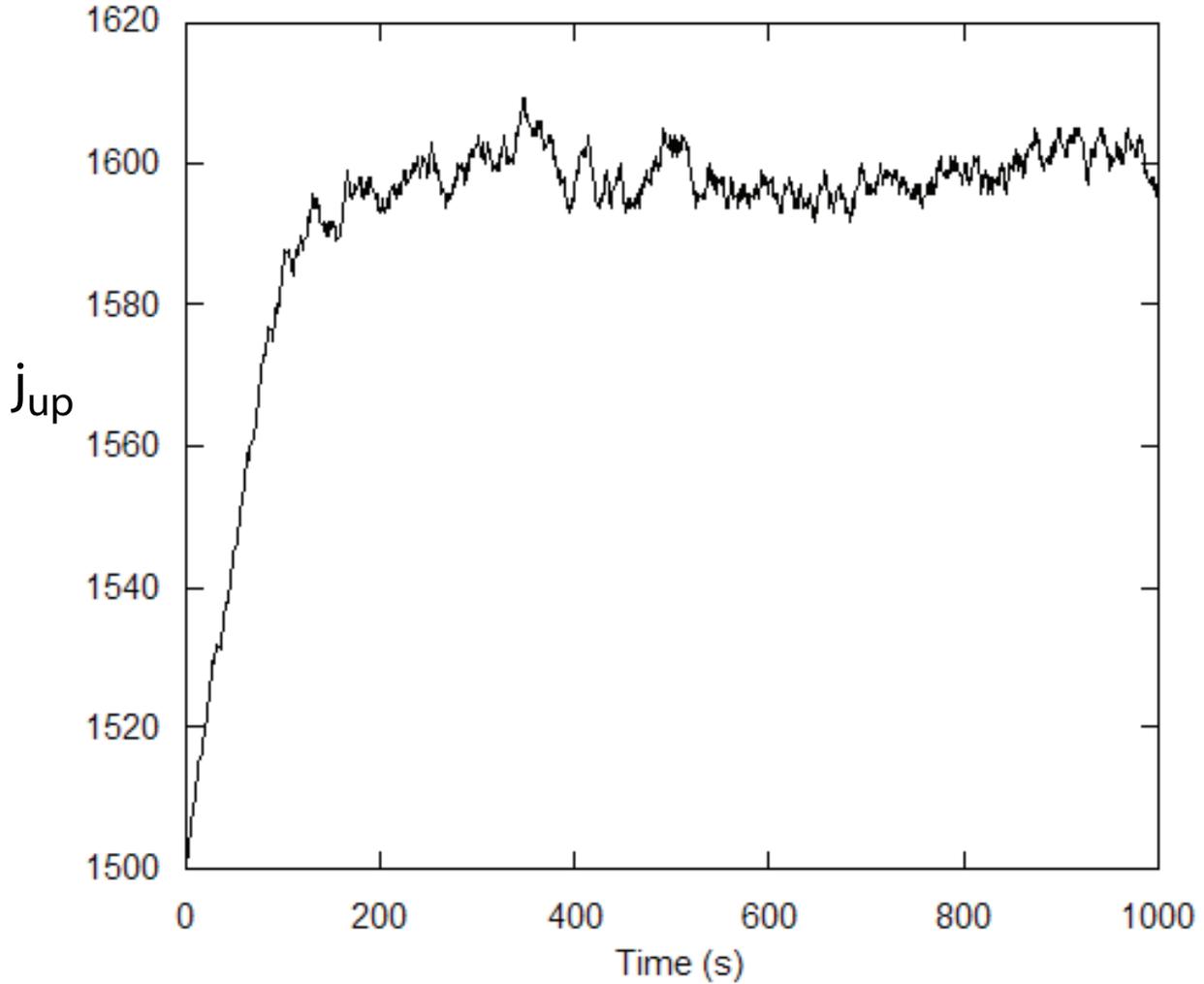}%
\caption{For the simulation described here, subsystem 2 has 3000 particles, so that 3001 macrostates are available.  With no external magnetic field the equilibrium value is 1500.  Because of the prescribed variation in subsystem 1 entropy production, the variable $j_2$ is pushed upwards away from equilibrium to a value near 1600.}%
\label{FigSim}%
\end{figure}

Equation~(\ref{expS}) has the attractive feature of simplicity; though analytically difficult to deal with without making simplifying approximations, it is readily amenable for numerical simulation.  It is quite straightforward to numerically simulate outcomes for one slow variable (\#1) coupled to one fast variable (\#2).  In the simulations presented here, very slow variable 1 is out of equilibrium and is producing entropy at a rate $\sigma_1$ while relaxing towards equilibrium.  This rate of entropy production depends on the value of variable 2.  We make system 2 very simple: $N_2$ non-interacting particles, each residing in one of two states (two level system such as the spin 1/2 paramagnet~\cite{Schroeder,Reif}).  The state of subsystem 2 is described by one discrete variable $j_2$ which is the number of particles with spin up.  This variable may take integer values from 0 to $N_2$.  In a zero magnetic field environment, the equilibrium value for $j_2$ would be $N_2/2$, ($N_2$ an even integer) if not coupled to system 1.  In our simulations we take time steps (one second each) during which we allow for system 2 to change $i_2$ value by one, either upwards or downwards.  The change in entropy during this time-step, due to subsystem 1, is specified by:
\begin{equation}
\Delta S_1=\sigma_{1_0}+c(j_2-N_2/2)        \label{eq:sim}
\end{equation}  
where $c$ is a constant describing the strength of the entropic induction effect.  Positive values of $c$ give a statistical preference for $j_2$ values above the equilibrium value.  The key step is to numerically calculate weighting factors $e^{\Delta S_1/k_B}$ for each possible outcome.  These important weighting factors are what gives the statistical preference.  During each time step more microstates are created and sampled (hence more weighting) if $\Delta S_1$ is larger.   Implementing Eq.~(\ref{eq:sim}) into Eq.~(\ref{expS}), along with a standard relaxation term for system 2 allows for calculation of the probability of subsystem 2 either making a transition upwards or downwards.  In Fig.~\ref{FigSim} we present a simulation in which subsystem 2 contains $N_2=3000$ particles, and begins in its equilibrium state.  As we can see the value of $j_2$ is pushed away from equilibrium, completely as a result of simple statistics.  Variable $j_2$ attains a new mean value.  Also visible are fluctuations, both in magnitude and in timescale $(\tau_2)$, in variable $j_2$, which are consistent with the strength of the dissipation constant trying to push subsystem 2 towards equilibrium. 

\section{Simple physical example: thermal conduction} \label{sec:thermal}

\begin{figure}[ht]%
\includegraphics[width=0.8\columnwidth]{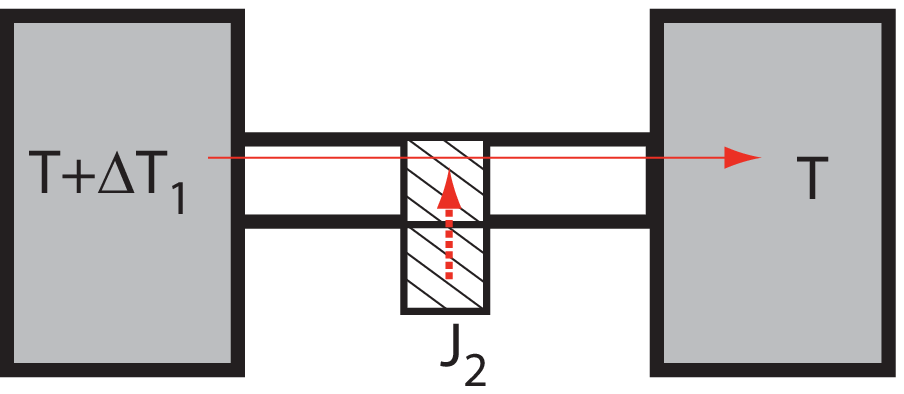}%
\caption{Schematic diagram for two variable thermal transfer system.  Shaded areas represent subsystem 1 with large heat capacity (slow dynamics), while the small hatched rectangle represents subsystem 2, a bottleneck composed of material having a significant temperature variance in the thermal conductivity.  Arrows denote fluxes in both subsystems. }%
\label{FigThermTrans}%
\end{figure}

We consider the simple two variable case where both variables represent energy imbalance i.e. $x_1=U_1$ and $x_2=U_2$.  So,
\begin{equation}
X_1=\Delta \left(\frac{\partial S}{\partial U_{1}}\right)=\Delta\frac{1}{T_1}=-\frac{1}{T_1^2}\Delta T_1   ~.
\end{equation}
Similarly $X_2=-\frac{1}{T_2^2}\Delta T_2$, and we identify the conjugate forces with temperature differentials~\cite{degroot}.  A schematic for this system is provided in Fig.~\ref{FigThermTrans}, which also illustrates how subsystem 2 can act like a bottleneck for the heat transfer in system 1.  The heat capacities $C_1$ and $C_2$ can be used as: $a_1=C_1\Delta T_1$, $a_2=C_2\Delta T_2$.
% note $X_1=-g_{11}U_1$ so $U_1=-X_1/g_{11}=\frac{1}{g_{11}T_1^2}\Delta T_1 $ and we see that $C_1=\frac{1}{g_{11}T_1^2}$
We note that $g_{11}=1/(C_1 T_1^2)$, $g_{22}=1/(C_2 T_2^2)$.  To linear order: $\dot{a}_1=L_{11}X_1$ i.e.  $\dot{U}_1=-L_{11}\frac{1}{T_1^2}\Delta T_1=-k_1\Delta T_1$, where $k_1\equiv L_{11}/T_1^2$, and we see that $L_{11}$ is proportional to the thermal conductivity coefficient.  Also $\dot{U}_1=-L_{11}\frac{1}{C_1 T_1^2}\Delta U_1$ which allows one to identify the timescale $\tau_1=\frac{C_1 T_1^2}{L_{11}}$.
%Now if $\lambda$ is the thermal conductivity then $J_1=-\lambda_1 (A_1/l_1)\Delta T_1$
%Comparing gives: $L_{11}=\lambda_1 (A_1/l_1)T_1^2$

We assume that the thermal conductivity coefficient $\lambda_1$ depends to some degree on temperature, so therefore on $\Delta T_2$.  If we define a dimensionless constant $\kappa\equiv\frac{T_2}{\lambda_1}\frac{\partial \lambda_{1}}{\partial T_2}|_{a_2=0}$ , then $\kappa\equiv\frac{T_2}{L_{11}}\frac{\partial M_{11}}{\partial T_2}|_{a_2=0}$
% $\frac{\partial L_{11}}{\partial T_2}|_{a_2=0}=\frac{A_1 T_1^2 \lambda_1}{l_1 T_2 } \kappa$
and $\gamma_{11,2}=\frac{\partial M_{11}}{\partial U_2}|_{\Delta T_2=0}= \kappa L_{11}T_2 g_{22}$.

%$\gamma_{11,2}=\frac{\partial L_{11}}{\partial U_2}|_{\Delta T_2=0}=\frac{\kappa \lambda_1 A_1 T_1^2}{C_2 l_1 T_2}=\frac{\kappa L_{11}}{C_2 T_2}= \kappa L_{11}T_2 g_{22}$
When $a_2$ is in a stationary state, then from Eq.~(\ref{X2stat}):
%\begin{equation}
%X_2 =- \frac{r_2\gamma_{11,2}}{g_{22}L_{22}} X_1^2 =-  \frac{r_2\kappa L_{11}T_2 }{L_{22}} X_1^2 =-\frac{r_2\kappa g_{22}\tau_2 T_2 }{g_{11}\tau_1} X_1^2 =-\frac{\kappa C_1 T_1^2\tau_2^*}{C_2 T_2\tau_1} X_1^2~, \label{X2example}
%\end{equation}
\begin{equation}
X_2 =- \frac{r_2\gamma_{11,2}}{g_{22}L_{22}} X_1^2 =- \frac{\tau_2^*\kappa L_{11}T_2 }{\tau_2 L_{22}} X_1^2 ~. \label{X2example}
\end{equation}
In terms of temperature differentials, assuming $T_1\approx T_2\equiv T$, and using $L_{22}g_{22}\tau_2=1$:
%\begin{equation}
%\Delta T_2 =  \frac{r_2\kappa k_1 T^2 T_2 }{k_2 T^2} \frac{(\Delta T_1)^2}{T^2} = \frac{\tau_2^*\kappa k_1  T_2 }{\tau_2 k_2 } \frac{(\Delta T_1)^2}{T^2} 
%= \frac{\tau_2^*\kappa k_1  T g_{22}L_{22} }{k_2 } \frac{(\Delta T_1)^2}{T^2} = \frac{\tau_2^*\kappa k_1  T g_{22}T^2 }{1 } \frac{(\Delta T_1)^2}{T^2}~, 
%\end{equation}
\begin{equation}
\frac{\Delta T_2}{T} = \tau_2^*\kappa \frac{ k_1 } {C_2 } \frac{(\Delta T_1)^2}{T^2}~. 
\end{equation}

If the characteristic lengthscale of the bottleneck region is $l$, then one can show that $\frac{k_1}{C_2}=\frac{\lambda_1}{c_2 l^2}$ where $\lambda_1$ is the material thermal conductivity of subsystem 1, while $c_2$ is the volumetric specific heat of subsystem 2.  Thus,
\begin{equation}
\frac{\Delta T_2}{T} = \tau_2^*\kappa \frac{\lambda_1}{c_2 l^2} \frac{(\Delta T_1)^2}{T^2}~. 
\end{equation}
For copper at room temperature, $\frac{\lambda_1}{c_2}=1.1\times 10^{-4}$ m$^2/$s~\cite{AM}. 
% If we take $l=32$ nm, which is currently a standard minimum feature size in lithographicly fabricated features in electronic technology, then $\frac{\lambda_1}{c_2 l^2}=1.1\times 10^{11}$ s$^{-1}$. 
If we take $l=10$ nm, then $\frac{\lambda_1}{c_2 l^2}=1.1\times 10^{12}$ s$^{-1}$.  If we take $\tau_2^*=1.5\times 10^{-13}$ s, which is the characteristic time scale for atomic vibrations and fluctuations~\cite{AM,Zangwill}~then $\tau_2^*\frac{\lambda_1}{c_2 l^2}\approx~0.2$.  For copper, the temperature variation of the thermal conductivity is rather small~\cite{AM}: at room temperature $\kappa=-300\times ~0.0039=-1.2$.  So
\begin{equation}
\frac{\Delta T_2}{T} \approx -0.24 \frac{(\Delta T_1)^2}{T^2}~. \label{copper}
\end{equation}
Thus, even for an ordinary material such as copper, induction effects could be observed for very small systems.  For example if $\frac{(\Delta T_1)}{T}=0.1$ then we predict that $\frac{\Delta T_2}{T} =- 0.0024$.
 
If however, the material in the junction is near a metal insulator transition, then the thermal conductivity can also see rapid changes with temperature, possibly giving very large values for $\kappa$.  Examples of such systems include Fe$_3$O$_4$ with a transition temperature near 122 K and BaVS$_3$ with a transition near 70 K~\cite{Tokura1998}.  With these types of materials $\Delta T_2$ may be increased from the expression in Eq.~(\ref{copper}) for copper by an order of magnitude or even more.  Fabrication of structures on length scales of 10 nm is currently not easy it may become accessable with technology in the near future.  Integrated circuit structures on the order of 30 nm are currently being produced on a wide scale.  We note the interesting possibility of effectively shifting the transition temperature of a material through the application of a generalized force associated with another subsystem variable.  Fabrication of very small structures capable of producing significant values of $\Delta T_2$ could have important applications in microelectronics.  Entropic induction could be used to cool small regions of an integrated circuit.  Such a type of cooling could provide an alternative to cooling using the Peltier effect.

Note that the induction effect may become prominent in systems that are not microscopic, for example models for traffic flow~\cite{Kerner2013}, as an example where small changes in certain parameters can create a bottleneck effect.  Further examples of test systems may be found by considering systems where particle number, not energy, is out of equilibrium.  The generalized force would be chemical potential difference, as opposed to temperature difference.  Also, we point out that the generalized force $X_1$ does not have to be created artificially.  For example very small systems exist naturally which are composed of atoms and molecules on the verge of chemical reaction.  In this case the generalized driving force is the affinity~\cite{degroot,Prigogine}.  In these very small systems the time scale $\tau_2$ can very easily be almost as small as $\tau_2^*$.  Thus, in these systems the induction effect can likely be very significant.

We point out that since the induction effect does not even exist to linear order, then the results presented here represent the leading term in the response (for fast variables).  Thus, even if $\kappa$ and the driving force $X_1$ are not small and the accuracy of the theory presented here (Eq.~(\ref{X2example})) is not high, it still represents a good starting point and the best estimate currently available.

\section{Conclusions}

We have developed a nonequilibrium thermodynamic theory that demonstrates an induction effect of a statistical nature.  We have shown that this thermodynamic induction can arise in systems that are naturally nonlinear through having non-constant kinetic coefficients, i.e. the VKC class.   In particular if a kinetic coefficient associated with a given thermodynamic variable depends on another, faster, variable then we have derived an expression that can predict the extent of the induction.  The induction is proportional to the square of the driving force.  The nature of the inter-variable coupling for the induction effect has similarities with the Onsager symmetry relations, though there is an important sign difference as well as the magnitudes not being equal.  We have found that the nonlinear effects from the induction can enhance the stability of stationary states, as the system approaches equilibrium.  The induction effect gives an entropy production rate term that opposes dissipation, which we refer to as the induced rate of entropy production.  The key step in identifying the thermodynamic induction effect was in indentifying certain variables to act as the dynamical reservoir.  At least one such, slow, variable is essential in the analysis.  The dynamical reservoir also plays a key role in arriving at a new nonequilibrium version of Le Chatelier's principle.

%Though we suspect that induction effects can be found in systems with variables that cannot be easily divided into fast and slow types, we do not yet have analytic expressions for this case.  

We have also developed a variational approach, based on optimizing entropy production.  On the question of resolving whether entropy production is minimized or maximized, we conclude that, at least for the nonlinear systems considered here, it is the free entropy production that is maximized.  The maximization occurs while the fast variables are quasistationary.  Thus, the stationary states of Prigogine, introduced in the context of the minimum entropy production principle, are still very useful, at least within the VKC class of systems.  The proof we have provided is simple and provides predictive power such as in establishing the values of the Legendre coefficients, as well as prescribing just how much faster the entire system approaches equilibrium in the quasistationary states.  Such predictive power has been absent in previous discussions of the maximum entropy production principle~\cite{Seleznev2006}.  The maximum entropy production principle can be expressed in various ways, depending on whether one wants to focus on the free entropy production, the slow variable rate of entropy production, or the total rate of entropy production.
%: one way directly on the free entropy production and two ways on either the total entropy production or the slow variable entropy production, with added contraints.

We anticipate that in some systems, thermodynamic induction effects are not merely small corrections to a linear response, but that the effects may be very significant.  We have shown that there exist non-equilibrium quantities analogous to the free energies of equilibrium thermodynamics.  These newly defined free entropies, which are non-zero only in non-equilibrium conditions, can quantify the significance of the induction effects.

Finally, we have discussed some schemes directed towards discovering experimental evidence for entropic induction, including a possible application to specialized cooling in integrated circuits.  Detailed calculations show that the entropic induction effect is most likely to be realized if a key component of the system is very small.  Inside such a small region, or junction, fluctuations play an important role behind the induction. 

%One should ask the question that if the forces are not small i.e. very far from equilibrium, the effects could be large and this would therefore be the best starting point to model such systems.
%The observable effects of entropic induction are likely to be small in most systems and one may have to look very carefully to see the effect.
%How to extend this work to induction of slow variables i.e. how effectively can a slow variable induce a fast one? (other way?)
%(first discussion of a free variable in non-equilibrium thermodynamics, for conclusions)
%\vspace{0.1in}

\bibliography{EntMan16Nov2013}

\end{document}